\documentclass{aa}  

\usepackage{graphicx}
\usepackage{txfonts}
\usepackage{hyperref}

\begin{document}

   \title{Finding the mechanism of wave energy flux damping in solar pores using numerical simulations}

   \author{J.M. Riedl
          \inst{1}
          \and
          C.A. Gilchrist-Millar
          \inst{2}
          \and
          T. Van Doorsselaere
          \inst{1}
          \and
          D.B. Jess
          \inst{2}
          \and
          S.D.T. Grant
          \inst{2}
          }

   \institute{Centre for mathematical Plasma Astrophysics (CmPA), KU Leuven \\ 
                Celestijnenlaan 200B bus 2400, 3001 Leuven, Belgium \\
                \email{juliamaria.riedl@kuleuven.be}
         \and
             Astrophysics Research Centre, School of Mathematics and Physics, Queen's University Belfast\\
             Belfast, BT7 1NN, U.K. }

   \date{Received; accepted}

 
  \abstract
  {Solar magnetic pores are, due to their concentrated magnetic fields, suitable guides for magnetoacoustic waves. Recent observations have shown that propagating energy flux in pores is subject to strong damping with height; however,  the reason is still  unclear.}
  {We investigate possible damping mechanisms numerically to explain the observations.}
  {We performed 2D numerical magnetohydrodynamic (MHD) simulations, starting from an equilibrium model of a single pore inspired by the observed properties. Energy was inserted into the bottom of the domain via different vertical drivers with a period of 30s. Simulations were performed with both ideal MHD and non-ideal effects.}
  {While the analysis of the energy flux for ideal and non-ideal MHD simulations with a 
plane driver cannot reproduce the observed damping, the numerically 
predicted damping for a localized driver closely corresponds with 
the observations. The strong damping in simulations with localized driver was caused by two geometric effects,  geometric spreading due to diverging field lines and  lateral wave leakage.}
  {}


   \maketitle
%

\defcitealias{gilchrist_etal_2021}{GM21}


\section{Introduction} \label{sec:introduction}

Solar pores are macroscopic features resembling small sunspots lacking a penumbra, but can also occur as a precursor or remnant of sunspots \citep{garciadelarosa_1987,sobotka_2003,thomas_weiss_2004}. Given their nearly circular symmetry and high magnetic field concentrations, pores act as efficient wave guides for magnetoacoustic waves, allowing wave flux to enter higher regions of the solar atmosphere \citep{jess_etal_2015} where the energy can then be dissipated \citep{grant_etal_2018}. 

The observational evidence of waves in solar pores is vast. Photospheric sausage modes in pores were identified by, e.g., \citet{fujimura_tsuneta_2009} (sausage and/or kink waves), \citet{morton_etal_2011} \citep[being fast waves according to][]{moreels_etal_2013}, \citet{dorotovic_etal_2014} (standing slow and fast waves), \citet{grant_etal_2015} (propagating slow surface waves), \citet{keys_etal_2018} (surface and body waves), and \citet{gilchrist_etal_2021} \citepalias[propagating slow surface and body waves, hereafter referred to as][]{gilchrist_etal_2021}. These authors all found evidence of wave periods of around 3 and/or 5 minutes, indicating the likely role of photospheric p-modes as a driver for the waves.

The propagation to the chromosphere was studied by \citet{balthasar_etal_2000}, who confirmed, by using the Vacuum Tower Telescope (VTT) on Tenerife, the presence of five-minute oscillations for the magnetic field in the deep photosphere, as seen in other observations. Using the Transition Region and Coronal Explorer (TRACE) observations, they found a peak at a period of three minutes in the chromosphere. \citet{stangalini_etal_2011} reported longitudinal acoustic waves reaching the chromosphere in both three- and five-minute bands. They underline the strong connection between wave transmission and magnetic field geometry, which suggests that for pore models special attention should be paid to the definition of the magnetic field, as also suggested by \citet{jess_etal_2013}. 

However, how far waves in solar pores propagate into higher layers of the solar atmosphere is still unclear. \citet{khomenko_collados_2006} conducted numerical simulations of waves in a small sunspot;  they used a localized wave source to study wave propagation, refraction, and mode conversion. They found that due to the vertical and horizontal stratification of the Alfv\'{e}n speed, (low $\beta$) fast waves are refracted in the chromosphere back down to the photosphere, while slow modes continue propagating up. 
Recent simulations of a chromospheric resonance layer above a sunspot done by \citet{felipe_etal_2020} show that actual wave propagation only takes place between the photosphere and chromosphere. A chromospheric resonance layer was previously also simulated by \citet{botha_etal_2011} and observed by \citet{jess_etal_2020}.
On the other hand, \citet{riedl_etal_2019} showed  in concentrated flux tubes that plane waves are converted to tube (sausage and kink) waves that are able to propagate to the corona since the tube structure greatly affects the wave propagation \citep{cally_khomenko_2019_part_I,khomenko_cally_2019_part_II}.

\citet{grant_etal_2015}, and more recently \citetalias{gilchrist_etal_2021}, measured wave energy throughout the lower atmosphere of solar pores, and indeed report significant energy flux damping as a function of height. Analytic calculations of \citet{yu_etal_2017} show that the observed damping could at least be partly explained by resonant damping of slow sausage waves. Although they find that this damping mechanism is stronger than previously expected, the numerical studies of \citet{chen_etal_2018}, validated by analytic calculations of \citet{geeraerts_etal_2020}, show that damping due to electrical resisitvity is much more potent than that due to resonant absorption. However, this alone is not enough to account for the damping. Flux could also be lost due to leaky tube waves \citep{cally_1986}. Leaking waves had already been observed by \citet{stangalini_etal_2011} and \citet{morton_etal_2012}. \citet{grant_etal_2015}  mentioned that part of the waves in their observations are reflected, which fits  the simulations of \citet{khomenko_collados_2006}, and that mode conversion \citep{cally_2001,bogdan_etal_2003} might play a role. Frequency dependent damping for slow magnetoacousic waves in sunspot umbrae is discussed by \citet{prasad_etal_2017}, who find that higher frequencies are damped more strongly. The authors suspect this behavior occurs due to radiative and/or conductive losses.

Another important factor to consider is the cutoff frequency  present in stratified media \citep{lamb_1909}. Acoustic waves with lower frequencies than the cutoff frequency cannot propagate, but are evanescent standing waves. This effect can be used to determine the cutoff frequency of the solar atmosphere \citep{felipe_etal_2018}, which indicates that five-minute waves like those observed by \citetalias{gilchrist_etal_2021} should be below the cutoff. However, the phase lag and propagation speed between different heights suggest that the observed waves in \citetalias{gilchrist_etal_2021} are indeed propagating as evanescent waves should not show any phase differences \citep{carlsson_stein_1997}. On the other hand,  the picture is not completely black and white. \citet{centeno_etal_2006} summarized the effects of the cutoff frequency;  when radiative losses are taken into account, they find that there is no clear value for the cutoff frequency, but a transition between mainly evanescent and mainly propagating waves. Therefore, it is possible that the waves in \citetalias{gilchrist_etal_2021} are partly subject to the cutoff, which could account for at least part of the observed damping. For the sake of this study, however, we assume the waves to be 100\% propagating.

In this paper we aim to expand our understanding of the damping mechanisms in solar pores by explaining the observed damping with simple two-dimensional (2D) numerical simulations, using a model inspired by the observational parameters obtained by \citetalias{gilchrist_etal_2021} for their pore 3. We insert propagating waves at the bottom of the domain with a vertical velocity driver with a frequency above the cutoff frequency, and study the resulting wave energy flux with height in comparison with the data from \citetalias{gilchrist_etal_2021} for different setups. In Section \ref{sec:methods} we briefly reiterate the most important points of \citetalias{gilchrist_etal_2021} before introducing the model, the numerical setup, and the approach for calculating the wave energy flux. The results, distinguished by driver location, are presented in Section \ref{sec:results} and thoroughly discussed in Section \ref{sec:discussion}. A short discussion about the case of a driver with frequency lower than the cutoff frequency is presented in Appendix \ref{appendix:7_min}.


\section{Methods} \label{sec:methods}

\subsection{Observations} \label{subsec:observations}

The model developed in this work is inspired by the observations detailed in \citetalias{gilchrist_etal_2021}, who utilized data obtained by the Facility Infrared Spectropolarimeter \citep[FIRS;][]{jaeggli_etal_2010} based at the National Science Foundation’s Dunn Solar Telescope (DST), Sacramento Peak, New Mexico. The FIRS data consist of sit-and-stare slit-based spectropolarimetric observations of the decaying active region NOAA 12564, which was captured between 14:09 - 15:59 UT on 2016 July 12 in the Si I 10827 \AA{} spectral line. The observations acquired contain a set of five solar pores that were positioned along a unique straight-line configuration. To cover all pores in a single FIRS exposure, the DST coude table was rotated so the spectrograph slit passed through the center of each photospheric pore boundary.

An examination of the photospheric Si I 10827 \AA{} line bisector velocities showed periods on the order of five minutes across all pore structures. Through spectropolarimetric inversions using the Stokes Inversion based on Response functions \citep[SIR;][]{ruizcobo_1992} code, the local plasma densities, magnetic field strengths, and temperatures were deduced as a function of atmospheric height spanning the range 0 – 500 km. The central pore \citepalias[pore 3 in][]{gilchrist_etal_2021} exhibited the best signal-to-noise ratio, and so was selected for comparison with the present theoretical work. 

\begin{figure}
  \centering
  \includegraphics[width=0.45\textwidth]{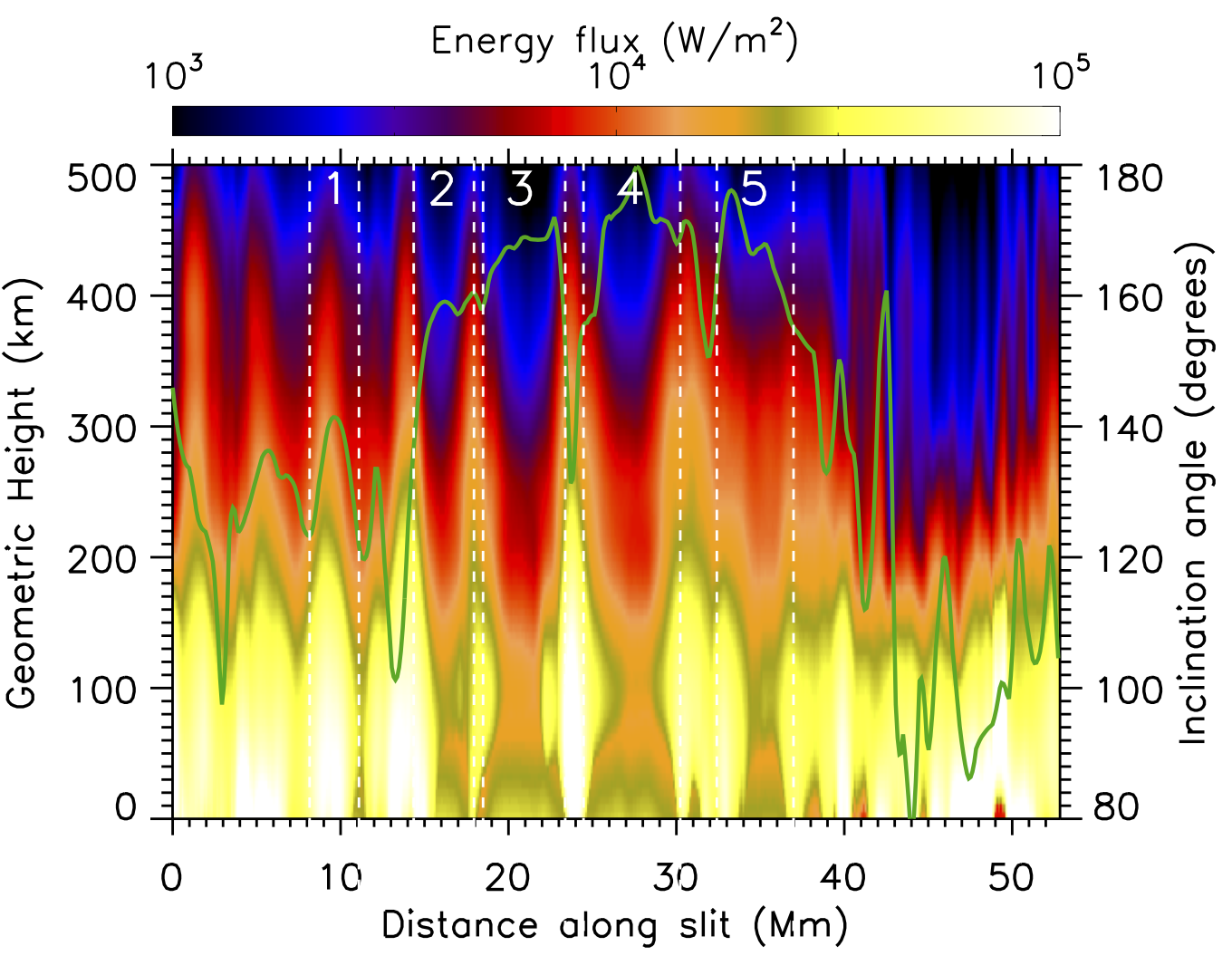}
  \caption{Energy flux across all five observed pores as a function of height. The color scale is logarithmic. Pore boundaries are shown as white dashed lines. The green solid  line shows the inclination angle of the magnetic field. From \citetalias{gilchrist_etal_2021}.}
  \label{fig:caitlin_energy_flux}%
\end{figure}

For pore 3 documented by \citetalias{gilchrist_etal_2021}, the magnetic fields were found to be close to vertical toward the pore center, with field strengths of ~2400 G and ~1000 G at atmospheric heights of 0 km and 500 km, respectively. Temperatures ranged from ~5000 K to ~3500 K and densities spanned from ~8.5x10$^{-4}$ kg m$^{-3}$ to ~9.8x10$^{-6}$ kg m$^{-3}$ across the same height range. Parameters derived from the inversions were combined with mean square velocities to calculate energy flux estimates as a function of atmospheric height (Equation \ref{eq:caitlin_flux}). The energy flux across all five pores as a function of height is displayed in Figure \ref{fig:caitlin_energy_flux}. Pore 3 was found to exhibit considerable energy damping with an average energy flux on the order of ~ 25 kW m$^{-2}$ at an atmospheric height of 100 km, dropping to ~ 1.5 kW m$^{-2}$ at 500 km. The damping mechanisms producing this drop in energy flux remain elusive. In addition, an increase in energy flux toward the boundaries of pore 3 indicated the presence of surface mode waves.

\subsection{Model} \label{subsec:model}

In order to investigate the wave damping in solar pores with numerical simulations, we first need to create a 2D gravitationally stratified magnetohydrostatic (MHS) equilibrium atmosphere that resembles the observational data. For a MHS equlibrium the following condition must be fulfilled
\begin{equation} \label{eq:MHS}
    \Vec{\nabla}p-\frac{1}{\mu_0}\left(\Vec{\nabla}\times\Vec{B}\right) \times \Vec{B}-\rho \Vec{g}=0,
\end{equation}
where $p$ is the gas pressure, $\rho$ is the density, $\Vec{B}$ is the magnetic field, $\mu_0$ is the magnetic permeability, and $\Vec{g}$ is the gravitational acceleration.

We start by choosing the magnetic field in the $z$-direction
\begin{equation} \label{eq:Bz}
    B_z(x,z)=a(z)\left[\arctan \left( \frac{x+r(z)}{s(z)}\right) - \arctan \left( \frac{x-r(z)}{s(z)}\right) \right] +b(z),
\end{equation}
which results in a bundle of strong vertical magnetic field of $a(z)$ inside the pore above the background field $b(z)$ outside the pore, resembling the observations. The parameter $r(z)$ describes the radius of the pore, while $s(z)$ is the smoothness parameter, which defines the thickness of the transition between pore and background. For the sake of simplicity, the written dependence on the vertical coordinate $(z)$ is   omitted from now on for these four parameters.

The parameters defining Equation \ref{eq:Bz} are 
\begin{equation}
    \begin{split}
        a & = & 0.33 G_{\mathrm{axis}} ~~~ [\mathrm{T}],~~~~~ r & = & \frac{2\times10^5}{\sqrt{G_{\mathrm{axis}}}} ~~~  [\mathrm{m}], \\
        b & = & 0.05 G_{\mathrm{side}} ~~~  [\mathrm{T}],~~~~~ s & = & 0.1r ~~~ [\mathrm{m}],
    \end{split}
\end{equation}
with exponential functions approximating the observed magnetic field strength of pore 3 from \citetalias{gilchrist_etal_2021} at the axis of the pore $G_{\mathrm{axis}}=0.1\exp (-z/300000)+0.07$ [T] and the side of the pore $G_{\mathrm{side}}=0.1\exp (-z/300000)+0.02$ [T].

Because $\mathrm{div}\Vec{B}=0$, we know that in 2D
\begin{equation}
    \frac{\partial B_x}{\partial x}= -\frac{\partial B_z}{\partial z}.
\end{equation}
Therefore, 
\begin{equation}
    \begin{aligned}
  B_x(x,z)  =  & -\int \frac{\partial B_z}{\partial z} dx +h(z) \\ 
            =  & -\frac{da}{dz}\left[ \frac{s}{2} \ln \left( \frac{(x-r)^2+s^2}{(x+r)^2+s^2} \right) \right. \\
             & \left. +(x+r) \arctan \left( \frac{x+r}{s}\right) -(x-r) \arctan \left( \frac{x-r}{s}\right) \right] \\ 
              & -a \left[ \frac{1}{2}\frac{ds}{dz}\ln \left( \frac{(x-r)^2+s^2}{(x+r)^2+s^2}\right)\right. \\
             & \left. +\frac{dr}{dz}\arctan \left( \frac{x+r}{s}\right) + \frac{dr}{dz} \arctan \left( \frac{x-r}{s} \right)   \right] -\frac{d b}{dz}x,
    \end{aligned}
\end{equation}
where we assume that $h(z)=0$ because then the solution is anti-symmetric around $x=0$.

In order to get a solution that fulfills both components of Equation \ref{eq:MHS},
\begin{equation} \label{eq:p-condition}
    \frac{\partial^2p}{\partial x \partial z}=\frac{\partial^2 p}{\partial z \partial x}
\end{equation}
must be true. By differentiating the $x$-component of Equation \ref{eq:MHS} with respect to $z$ and the $z$-component with respect to $x$, and combining the resulting derivatives with Equation \ref{eq:p-condition}, we find a constraint for the density,
\begin{equation} \label{eq:drho}
  \begin{aligned}
    \frac{\partial \rho}{\partial x}= \frac{1}{\mu_0 g} & \left[ \frac{\partial B_x}{\partial x}\left( \frac{\partial B_z}{\partial x}-\frac{\partial B_x}{\partial z}\right) + B_x \left( \frac{\partial^2 B_z}{\partial x^2}-\frac{\partial^2 B_x}{\partial z \partial x} \right) \right. \\
    & \left. + \frac{\partial B_z}{\partial z}\left( \frac{\partial B_z}{\partial x}-\frac{\partial B_x}{\partial z}\right) + B_z \left( \frac{\partial^2 B_z}{\partial x \partial z}-\frac{\partial^2 B_x}{\partial z^2} \right) \right],
  \end{aligned}
\end{equation}
assuming $\Vec{g}=[0,-g]$ with $g=274$ m/s. The density can then be obtained with
\begin{equation} \label{eq:rho}
    \rho(x,z)=\int \frac{\partial \rho}{\partial x} dx + f(z).
\end{equation}
The function $f(z)$ is of great importance here as it defines the gravitational stratification of the density. We therefore set $f(z)$ to be equal to the average density obtained from the observations of \citetalias{gilchrist_etal_2021} for their pore 3. Since $\partial \rho/\partial x$ also has a dependence on $z$, we add a small constant to $\rho$ to ensure its non-negativity. Due to the complexity of Equation \ref{eq:rho} it is solved numerically.

From the second component of Equation \ref{eq:MHS}, the pressure can be calculated with
\begin{equation} \label{eq:prs}
    p(x,z)=\int \frac{\partial p}{\partial z} dz +j(x).
\end{equation}
As long as the pressure is symmetric around the pore axis at $x=0$, there is no need to add a function $j(x)$. However, we add a constant to ensure a positive pressure. This equation is also solved numerically.

Theoretically, the described model is in MHS equilibrium. However, numerical calculations as used in the solution of the model and in the simulation code itself are imperfect, often resulting in somewhat unstable behavior, especially when gravity is involved. Therefore, using the boundary conditions described in Section \ref{subsec:numerics}, we simulate the model without driver for a physical time of 1300 seconds to let it settle down. After this time, there are no significant changes to density, magnetic field, or pressure on a timescale compared to a few driver periods. This slightly relaxed atmosphere is then used as the initial condition for our simulations. We note, however, that even after the slight relaxation there are still significant velocities within the domain, meaning that the resulting model atmosphere has not completely settled to a MHS equilibrium.

\begin{figure}
  \centering
  \includegraphics[width=0.45\textwidth]{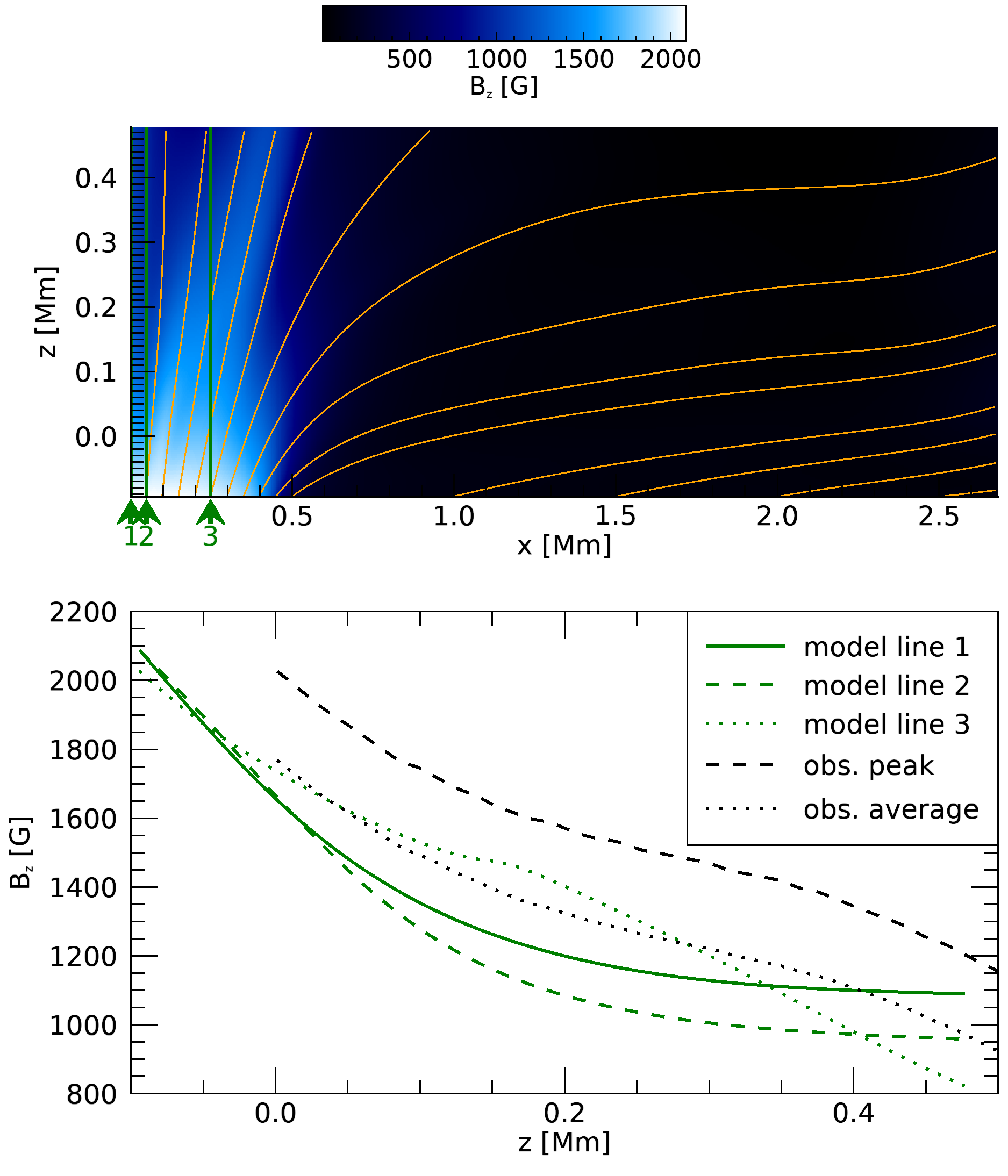}
  \caption{\textit{Top:} Vertical component of the magnetic field of the settled model atmosphere. Orange lines depict the magnetic magnetic field lines. \textit{Bottom:} Comparison of the model atmosphere with the observations from \citetalias{gilchrist_etal_2021} of pore 3. The maximum observational value within the pore is shown by \textit{obs. peak}, while the horizontal average across the whole pore is shown by \textit{obs. average}. The green lines show the model values for the indicated lines in the top figure (line 1 at pore axis).}
  \label{fig:magnetic_field}%
\end{figure}

\begin{figure}
  \centering
  \includegraphics[width=0.4\textwidth]{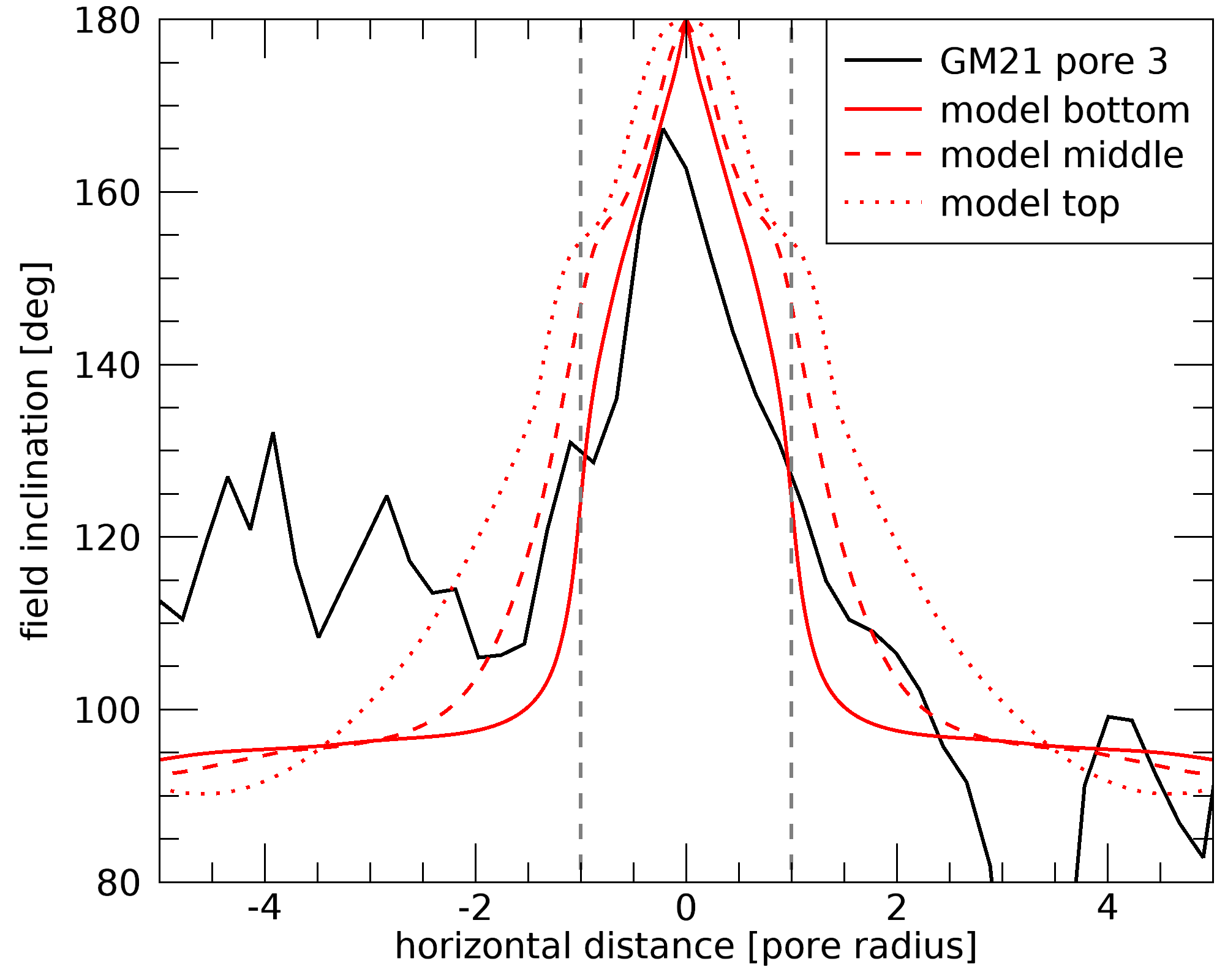}
  \caption{Comparison of the magnetic field inclination between model (red) and observations for pore 3 of \citetalias{gilchrist_etal_2021} (black) as a function of horizontal distance normalized to the pore radius. The pore radius for the model was assumed to be 0.44 Mm, while the radius of the observed pore is 2.5 Mm. The observational values are taken along the line perpendicular to the slit. Model values are mirrored around $x=0$ and are shown for the bottom (solid line), middle (dashed line), and top part (dotted line) of the model. The vertical dashed gray lines show the border of the pores at $x=1$.}
  \label{fig:field_inclination_comparison}%
\end{figure}

The top panel of Figure \ref{fig:magnetic_field} shows the vertical magnetic field component of the initial atmosphere, with field lines shown in orange. Due to the symmetry of the problem, only half of the pore is included in our model, with the pore axis being located at $x=0$. The pore itself is located on the left side of the plot, where the magnetic field is strong and mainly vertical. The deviation of the horizontal profile from the arctan-shape of Equation \ref{eq:Bz} occurs because of the equilibration process. The comparison of the model with the observations of \citetalias{gilchrist_etal_2021} (Figure \ref{fig:magnetic_field} bottom) shows great similarity. It should be noted, however, that the horizontal extent of our model pore (FWHM radius $\approx 0.44$ Mm) is smaller than the pores in the observations (radius $\approx 2.5$ Mm). Even so, when comparing the magnetic field inclination of the model atmosphere with the field inclination of pore 3 from \citetalias{gilchrist_etal_2021} in the direction perpendicular to the slit (thus perpendicular to the line of five pores), while taking the different radii into account, the field inclinations also coincide quite well (see Figure \ref{fig:field_inclination_comparison}). The plasma-$\beta$ in our model is higher than unity everywhere, with values ranging from 2 to 6.5 inside the pore and higher values up to 40 and higher outside.

\begin{figure}
  \centering
  \includegraphics[width=0.45\textwidth]{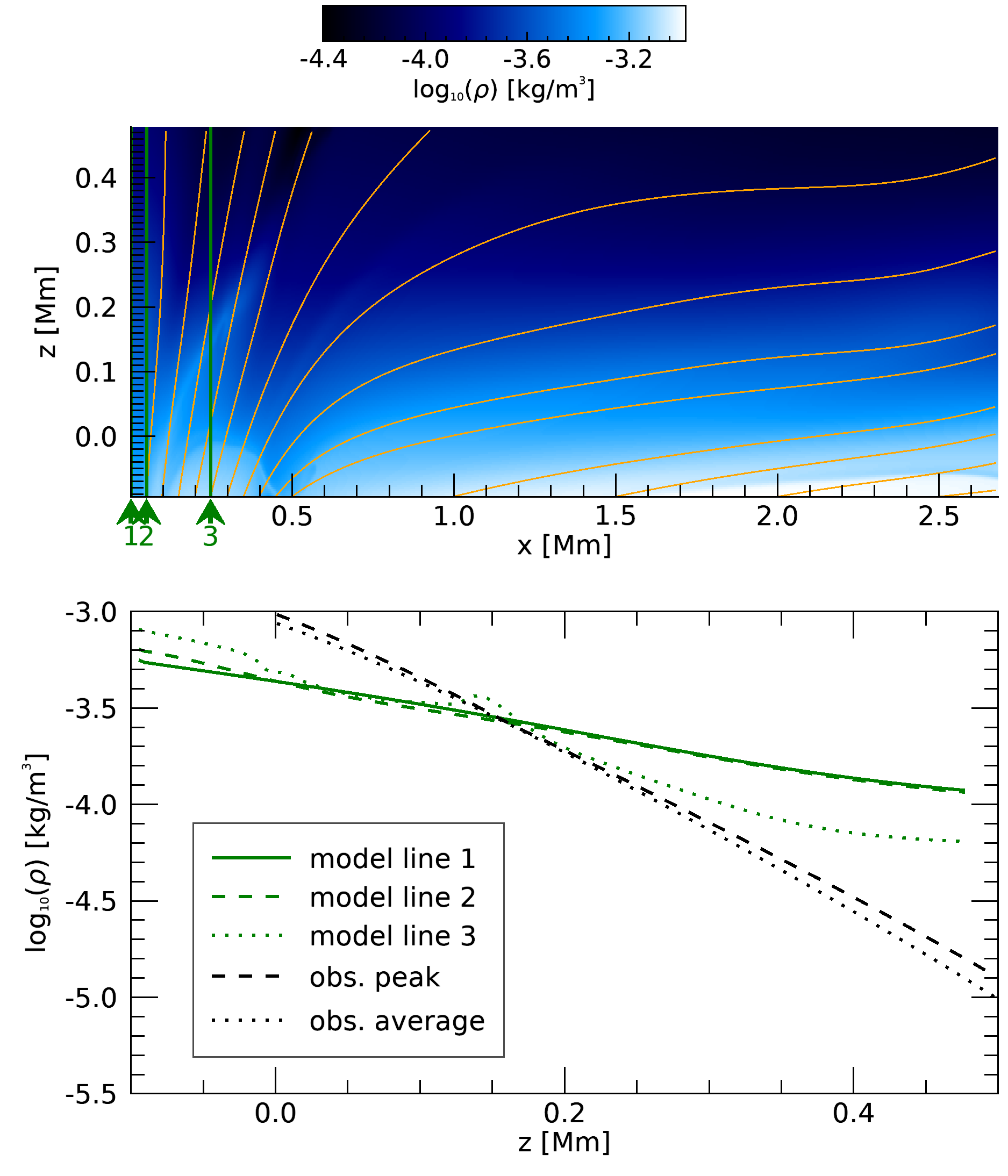}
  \caption{\textit{Top:} Logarithm of the density of the settled model atmosphere. Orange lines depict the magnetic magnetic field lines. \textit{Bottom:} Comparison of the model atmosphere with the observations from \citetalias{gilchrist_etal_2021} of pore 3. The maximum observational value within the pore is shown by \textit{obs. peak}, while the horizontal average across the whole pore is shown by \textit{obs. average}. The green lines show the model values for the indicated lines in the top figure (line 1 at pore axis).}
  \label{fig:density}%
\end{figure}

Similarly, Figure \ref{fig:density} shows the density of the settled model and the comparison to observations, where the density structures seen appear during the equilibration process. It is immediately apparent, that the model density is much less stratified with height than the observations, even though we added the observational density as stratification in Equation \ref{eq:rho}. This is caused by the effect of $\partial \rho / \partial x$ calculated by Equation \ref{eq:drho} already having a dependence on $z$, which in total decreases the stratification. In addition, it is also slightly decreased when  the atmosphere is allowed to settle down. However, it should be noted that for the observations in \citetalias{gilchrist_etal_2021}, the density is not a direct output of the inversions, but is instead determined through solving equations of state using inferred inversion outputs, under the assumption of hydrostatic (HS) equilibrium. This simplifying assumption is problematic in strong magnetic fields as it ignores the Lorentz force, thus providing notable uncertainties on the densities input into the model, of up to an order of magnitude \citep{borrero_etal_2019}. 

Nonetheless, the density values from \citetalias{gilchrist_etal_2021}   are still consistent with those from semi-empirical models like that of  \citet{maltby_etal_1986}, who considered a magnetized atmosphere at the center  of a sunspot umbra. They  also assumed a HS equilibrium; however, this assumption is valid for the center of an axially symmetric sunspot as the magnetic terms in Equation \ref{eq:MHS} vanish. Therefore, we have to assume that the observational values of the density are more reliable than the model values.

The smaller pore radius and less stratified density in our model compared to the observations are due to compromises being made when solving Equation \ref{eq:MHS}. Once a non-force-free magnetic field is chosen, the density or pressure cannot be freely chosen, but only manipulated through the addition of integration constants, as can be seen in Equations \ref{eq:rho} and \ref{eq:prs}. Therefore, in order to obtain a stable model for our simulations, certain concessions have to be made. In addition, due to the same reasons, our model also results in a plasma-$\beta>1$   inside the pore, as opposed to a low $\beta$ found by \citetalias{gilchrist_etal_2021} within the pores. The impact of the differences between observations and theoretical model on our results is discussed in Section \ref{subsec:differences_obs_model}.

\subsection{Numerical setup} \label{subsec:numerics}

All our simulations are conducted using the PLUTO code \citep{mignone_etal_2007}, which solves the magnetohydrodynamic (MHD) equations when using the respective module. Fluxes are computed using a linearized Roe Riemann solver, while the time step is advanced using an unsplit second-order accurate characteristic tracing method, which is less dissipative than multi-step algorithms. To deal with the inevitable occurrence of $\mathrm{div}\Vec{B}$ we use the mixed hyperbolic/parabolic divergence cleaning technique of \citet{dedner_etal_2002}, which is further discussed in \citet{mignone_etal_2010}. Gravity is added using a body force with constant acceleration toward the negative $z$-direction.

Keeping a model atmosphere stable when gravity is  included can often prove difficult and is highly contingent on the boundary conditions at boundaries perpendicular to the gravitational acceleration. In our case it was not possible to set fully open boundary conditions at the top boundary. We therefore expand the model atmosphere at the top to add a thick  high-viscosity layer to absorb all outgoing waves, effectively having an open boundary. We use the same boundary condition for the right boundary. The viscosity is treated with an explicit time integration. Including the viscous layers the domain ranges from 0 to 3 Mm in the $x$-direction  and from -0.095 to 0.795 Mm in the $z$-direction with $1000 \times 297$ cells, leading to a spatial resolution of 3 km in both directions. Excluding the viscous layers, a physical domain remains ranging from 0 to 2.68 Mm with 894 cells in the $x$-direction and from -0.095 to 0.475 Mm with 191 cells in the $z$-direction. Only this physical domain is used for the analysis and figures. The height of 0 Mm is defined as the bottom of the photosphere. After settling the calculated model from Section \ref{subsec:model} for 1300 s (defined as $t=0$ s in the plots), the simulations are run for an additional 200 s.

Due to the symmetry of the system our model only includes half of a solar pore, with the pore axis being located at the left boundary. Thus, the boundary conditions there are set to be reflective. At the bottom boundary we set pressure, density, and magnetic field to fixed values that fulfill the equations presented in Section \ref{subsec:model} for the initial model before the equilibration. The horizontal velocity is set to 0. For simulations without driver, the vertical velocity is set to 0 as well. When a driver is included the vertical velocity is set according to
\begin{equation} \label{eq:driver}
  v_{z,\mathrm{driver}}=A \sin\left(\frac{2 \pi}{T}t\right),
\end{equation}
with the amplitude $A=160$ m/s and the period $T=30$ s. Since the driver purely perturbs the velocity, some of the driver energy immediately flows into pressure and density perturbations. Between the ghost cells (additional cells outside the computation domain to enable numerical integration) including the driver and the first cell of the domain, the root mean square of the velocity perturbation is therefore reduced to levels observed by \citetalias{gilchrist_etal_2021} at the bottom of the pores of about 50 m/s. This short period for the driver was chosen because a typical p-mode period of 300 s is close to the cutoff period in our model, leading to the formation of standing waves due to reflections. However, we want to investigate propagating waves and their damping. In addition, for longer periods the wavelength would increase accordingly, causing the resulting waves to not fit into the domain. For the sake of completeness, we also did simulations with a low-frequency driver below the cutoff frequency, and we show a crude analysis in Appendix \ref{appendix:7_min}.

For some of our simulations we include non-ideal effects like viscosity, resistivity, and thermal conduction. Those effects were added using explicit time integration, and for expected values in the photosphere \citep[$R_e$ and $R_m$ taken from][]{ossendrijver_2003}. In the case of the simulations with viscosity, where viscosity was also present in the physical domain, simulations were done with exaggerated values for the viscous shear coefficient.

\subsection{Wave energy flux} \label{subsec:flux}

The energy flux can be calculated as \citep[e.g.,][]{goedbloed_poedts_2004}
\begin{equation} \label{eq:flux_basic}
    \Vec{E}=-\frac{1}{\mu_0}\left( \Vec{v}\times\Vec{B}\right)\times\Vec{B} + \left( \frac{\rho v^2}{2}+\rho \Phi + \frac{\gamma}{\gamma-1}p\right) \Vec{v},
\end{equation}
where $\Phi=-gz+\mathrm{const.}$ is the gravitational potential. The left term of Equation \ref{eq:flux_basic} is the Poynting flux, which is the magnetic component of the energy flux, whereas the other terms describe the hydrodynamic component. Since in our model $\beta>1$ everywhere and the driver mainly excites acoustic waves, the hydrodynamic component is dominant in our simulations.

There are still velocities up to nearly 2 km/s within the whole physical domain or up to $\sim$ 350 m/s within the pore after settling the atmosphere for 1300 s. These velocities are higher than the driver amplitude. Thus, in addition to simulations with a driver, we also conduct simulations without a driver, allowing us to  extract the effects caused by the input waves alone. This is done by subtracting all the variables of the simulations without a driver from the variables of the simulations with a driver, effectively giving us the perturbed variables
\begin{align*}
    \rho'=& \rho_\mathrm{driver}-\rho_\mathrm{no driver}, & p'=& p_\mathrm{driver}-p_\mathrm{no driver}, \\
    \Vec{v}'=& \Vec{v}_\mathrm{driver}-\Vec{v}_\mathrm{no driver}, & \Vec{B}'=& \Vec{B}_\mathrm{driver}-\Vec{B}_\mathrm{no driver}. 
\end{align*}
To obtain the wave energy flux, these perturbed variables are put into Equation \ref{eq:flux_basic}, in a process that is  similar to linearization.

In \citetalias{gilchrist_etal_2021}, on the other hand, the wave energy flux was calculated as
\begin{equation} \label{eq:caitlin_flux}
    E=\rho v_g \langle v^2 \rangle,
\end{equation}
with $v_g$ being the group speed and $\langle v^2 \rangle$ being the mean square velocity. For our simulations,   Equations \ref{eq:flux_basic} and \ref{eq:caitlin_flux} yield similar trends with absolute values in the same order of magnitude. Using Equation \ref{eq:flux_basic} facilitates a more detailed analysis, which is possible due to the much more detailed knowledge of the data in simulations compared to observations.


\section{Results} \label{sec:results}

We conducted a range of simulations,   including and removing non-ideal effects, and applying differing drivers.
Depending on the driver location, the results can be divided into two distinct groups, which are   discussed in the following.

\subsection{Driver located at whole bottom boundary} \label{subsec:full_driver}

\begin{figure}
  \centering
  \includegraphics[width=0.4\textwidth]{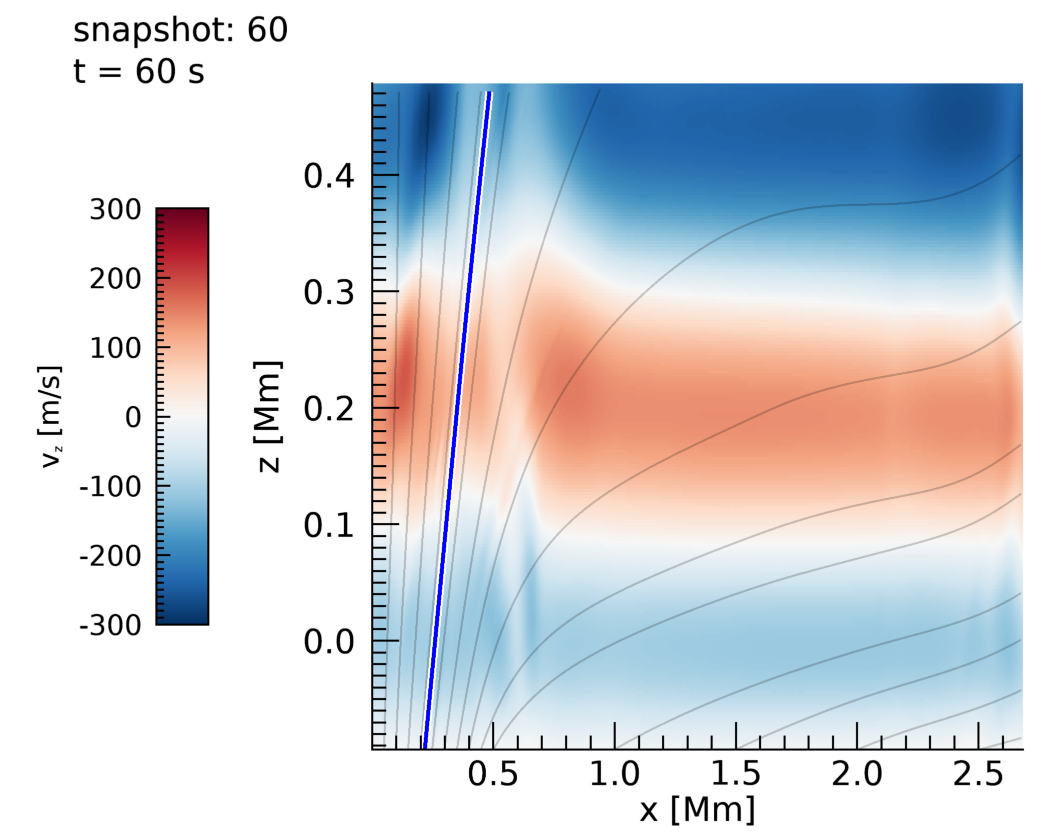}
  \caption{Snapshot of the vertical velocity perturbation after two periods for the full driver and ideal MHD. The gray lines show magnetic field lines. The blue line highlights the field line considered for the analysis in Figure \ref{fig:flux_full_driver}. The full time sequence is available as a movie online.}
  \label{fig:vz_full_driver}
\end{figure}

We applied the velocity driver described in Equation \ref{eq:driver} on the whole bottom boundary, resulting in plane fast waves propagating upward at approximately the sound speed. A single snapshot of the vertical velocity perturbation after two driver periods is shown in Figure \ref{fig:vz_full_driver}. The amplitude of the waves increases with height, as is expected due to the conservation of energy in a stratified plasma. The wave fronts are not completely horizontal, but have a jagged form at the pore location. This happens due to differing wave speeds at different locations. The vertical wave ridges visible at $x\approx 0.6$ Mm and the right boundary, and more pronounced at later times, as seen in the movie of the time sequence, are wave fronts of slow waves.

\begin{figure}
  \centering
  \includegraphics[width=0.4\textwidth]{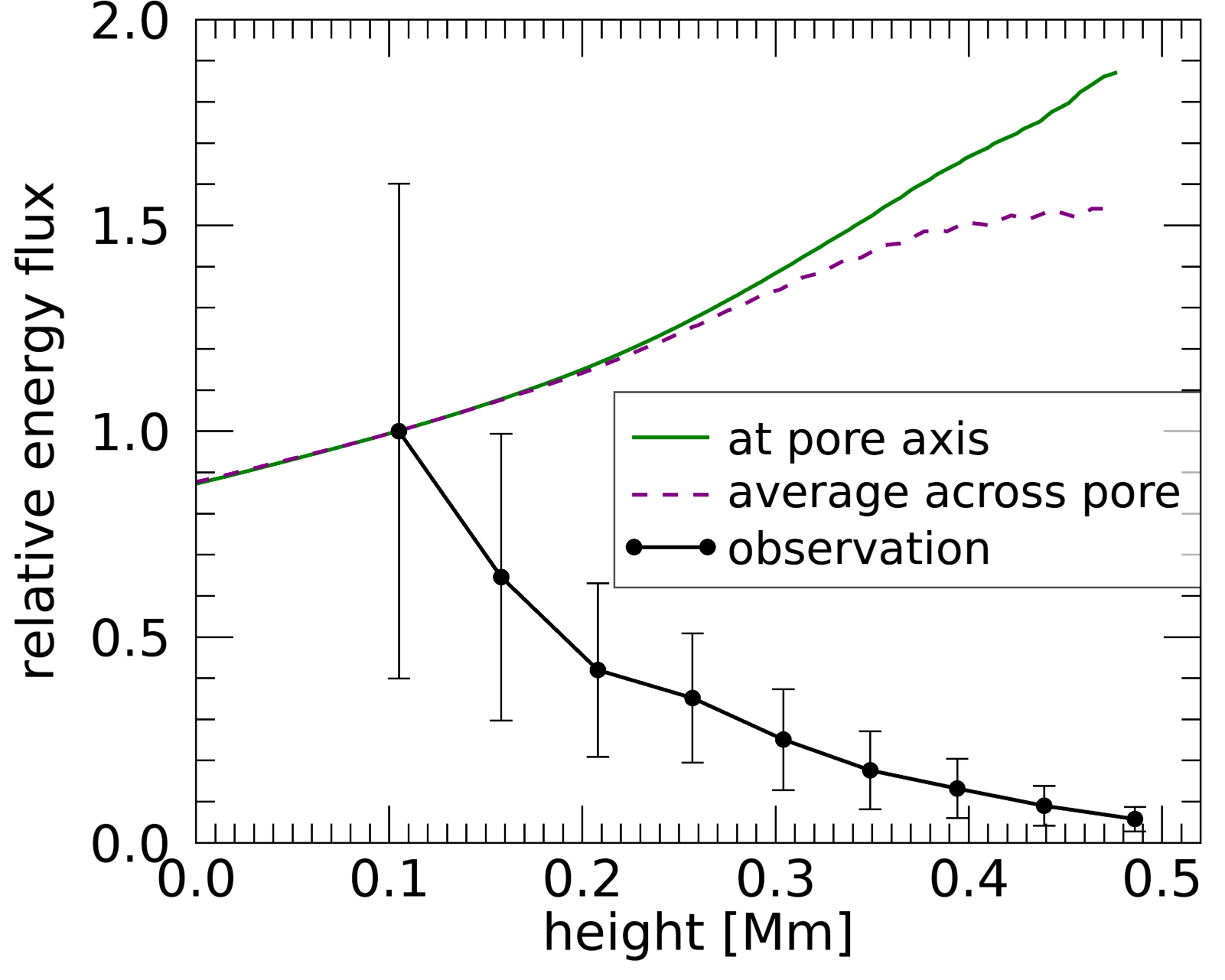}
  \caption{Relative wave energy flux parallel to the magnetic field averaged over time as a function of height for the full driver. The solid green line shows the energy flux along the pore axis, whereas the dashed purple line shows the average flux across the pore up to the field line highlighted in Figure \ref{fig:vz_full_driver}. The observational data (black line with symbols) are from pore 3 of \citetalias{gilchrist_etal_2021}. All fluxes are normalized to the first observational data point.}
  \label{fig:flux_full_driver}
\end{figure}

Figure \ref{fig:flux_full_driver} shows the time-averaged wave energy flux as a function of height relative to the first measure point obtained from the observations of \citetalias{gilchrist_etal_2021} for their pore 3. Both simulation and observational data were normalized to the data point at $z=0.1$ Mm. The time average of the simulation data was taken over the first period of the propagating wave. The figure shows the energy flux at the pore axis, where the magnetic field line is vertical (green line), and the flux averaged from the pore axis to the location of the field line highlighted in Figure \ref{fig:vz_full_driver} (purple dashed line). The observational data were also averaged in time and across the pore.

It is evident from Figure \ref{fig:flux_full_driver} that there is no indication of wave damping with height in our simulations; instead,  the energy flux   even increases with height, which could be explained by waves being refracted into the pore, as discussed in Section \ref{subsec:differences_obs_model}. The lack of damping is not only the case for ideal MHD, but also when resistivity, viscosity, or thermal conduction is included. Even for exaggerated viscosity no damping is achieved. We assume this is the case because we are studying a very narrow slab of atmosphere of a few hundred kilometers, leaving little time for non-ideal effects to affect the waves. Therefore, we fail to reproduce the observed damping with a plain driver located at the whole bottom boundary.

\subsection{Localized driver} \label{subsec:localized_driver}

Solar pores are magnetic structures that do not form in the photosphere but are already present below the solar surface. As solar pores are good wave guides, it is valid to assume that only the pore itself may be driven. Numerical simulations \citep{cameron_etal_2007} supported by observations \citep{cho_etal_2013} suggest that rapid cooling within pores could lead to downflows that collide with the plasma of lower layers to produce rebounding upflows, which further motivates the assumption of a localized driver. Moreover, previous simulations \citep{kato_etal_2016} show that photospheric buffeting by turbulent motions lead to the  efficient excitation of waves. We therefore alter our driver to a step-function driver that is only present in the inner part of the pore
\[ 
                v_{z,\mathrm{driver}}=
                \begin{cases}
                 A \sin\left(\frac{2\pi}{T}t\right) & x \le 0.2\mathrm{Mm} \\
                0 & x > 0.2\mathrm{Mm} .
                \end{cases}
\]

\begin{figure}
  \centering
  \includegraphics[width=0.4\textwidth]{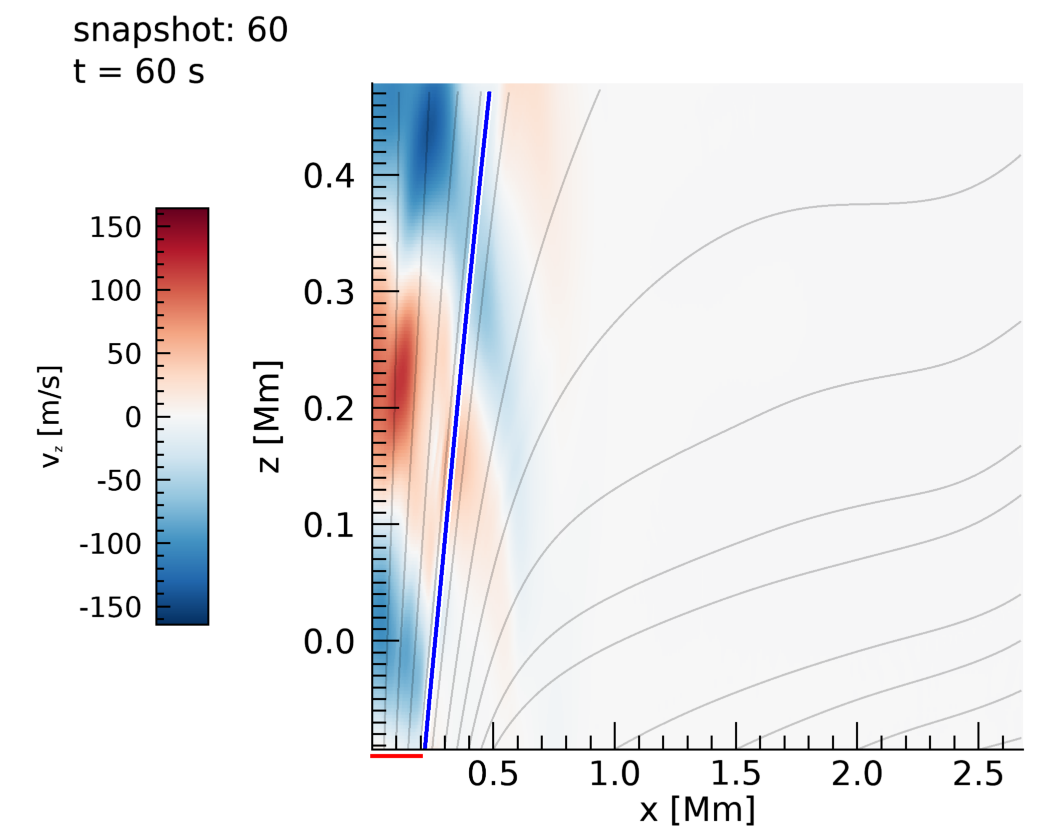}
  \caption{Snapshot of the vertical velocity perturbation after two periods for the localized driver for ideal MHD. The gray lines show magnetic field lines. The red bar below the $x$-axis indicates the driver location. The blue line highlights the field line considered for the analysis in Figure \ref{fig:flux_slab_driver}. The full time sequence is available as a movie online.}
  \label{fig:vz_slab_driver}
\end{figure}

\begin{figure}
  \centering
  \includegraphics[width=0.45\textwidth]{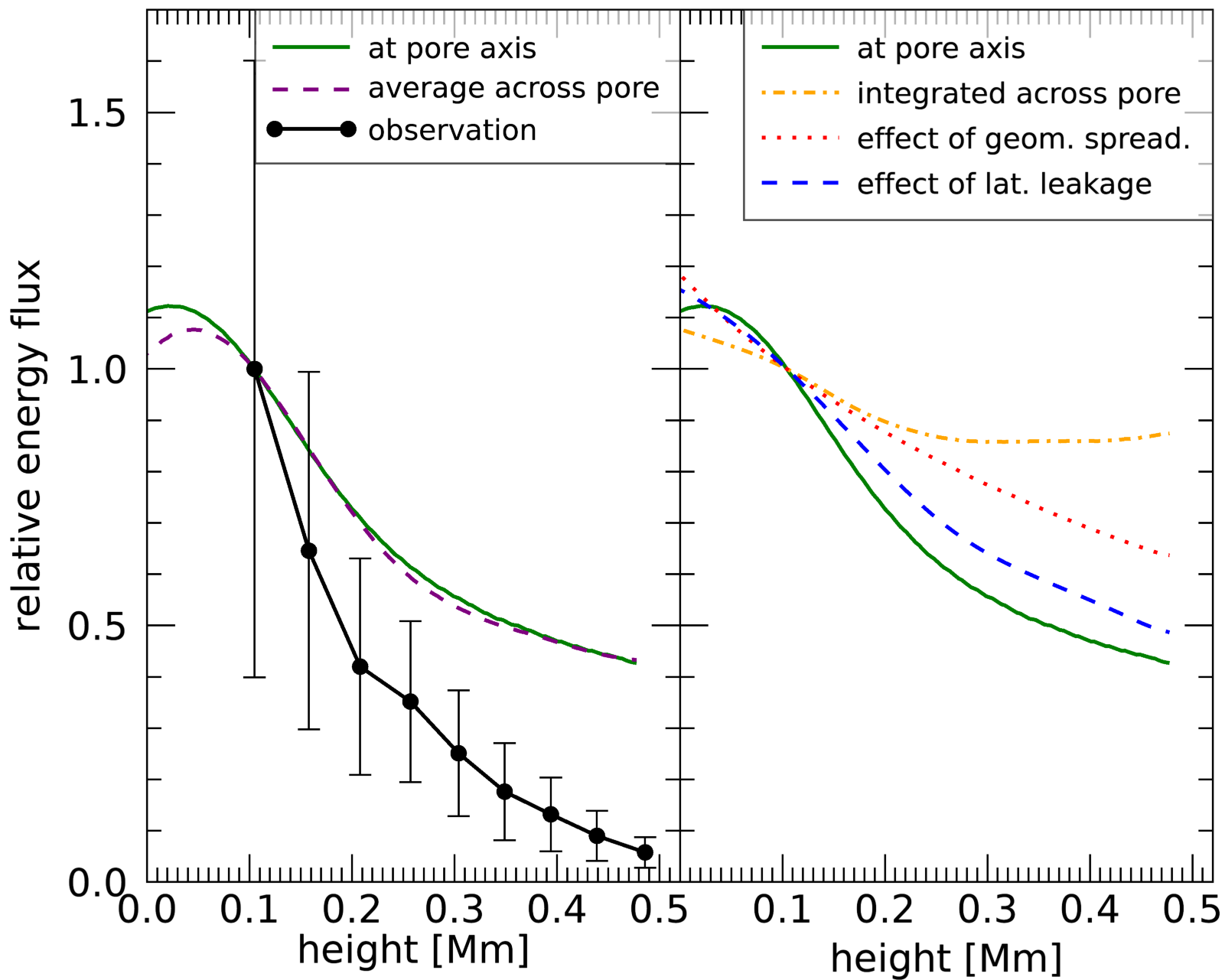}
  \caption{\textit{Left:} Relative wave energy flux parallel to the magnetic field averaged over time as a function of height for the localized driver. The solid green line shows the energy flux along the pore axis, whereas the dashed purple line shows the average flux across the pore up to the field line highlighted in Figure \ref{fig:vz_slab_driver}. The observational data (black line with symbols) are from pore 3 of \citetalias{gilchrist_etal_2021}. All fluxes are normalized to the first observational data point. \textit{Right:} Comparison of flux damping in the simulation with localized driver with the effects of geometric damping.  The solid green line shows the same data as in the left plot for comparison. The other lines are explained in the text.}
  \label{fig:flux_slab_driver}
\end{figure}

Figure \ref{fig:vz_slab_driver} shows a snapshot of the vertical velocity perturbation after two periods for the step-function driver. In contrast to the respective figure for the full driver, the wave fronts are not horizontal and the maximum amplitude is lower because the atmosphere is only driven at one location (indicated by the red bar below the $x$-axis). The blue line highlights a field line rooted slightly outside the driver location at $x=0.22$ Mm. There are clearly waves present beyond this field line, suggesting that the waves do not purely propagate along the magnetic field.

If we now study the wave energy flux as a function of height for the simulation with localized driver, as shown in Figure \ref{fig:flux_slab_driver} (left), it is immediately apparent that the energy flux is now strongly damped, in stark contrast to the simulations with the full driver. This sudden drop in wave energy flux with height by just changing the driver location can be explained by two geometric mechanisms: geometric spreading and lateral wave leakage. 

\subsubsection{Geometric spreading} \label{subsubsec:geometric_spreading}

The magnetic field lines in our model diverge with height. Therefore, if the waves were perfectly propagating along the field lines, the flux along a single field line, as well as the average flux at each height within the pore, would be expected to drop due to the flux being distributed across a wider area with increasing heights. The decrease in flux with height due to this mechanism is proportional to $1/R$ in 2D geometry, where $R$ is the distance between the pore axis and a specific field line. Such a curve is shown in Figure \ref{fig:flux_slab_driver} (right,  dotted red line) for the field line highlighted in Figure \ref{fig:vz_slab_driver}. Since this curve drops substantially less with height than the wave flux, there must be another mechanism with approximately equal significance. 

In addition, if only geometric spreading  caused the damping, the wave flux parallel to the magnetic field integrated across the pore should be constant with height because the same total amount of flux would be contained inside the pore at all heights. This is not the case, which can be seen with the dash-dotted orange line in Figure \ref{fig:flux_slab_driver}. Therefore, flux must escape from the pore through its edges.

\subsubsection{Lateral leakage} \label{subsubsec:lateral_leakage}

In our simulations with a localized driver, we observe waves propagating out of the solar pore, which decreases the flux inside the pore. This is the case because magnetoacoustic waves can propagate at an inclined angle with respect to the magnetic field. In a homogeneous plasma, the phase speed of fast and slow magnetoacoustic waves is \citep[e.g.,][]{goedbloed_etal_2019}
\begin{equation} \label{eq:phasespeed}
    v_\mathrm{fa/sl}(\theta)=\frac{\sqrt{v_s^2+v_A^2}}{\sqrt{2}} \left[ 1 \pm \left( 1-\frac{4v_c^2\cos^2 \theta}{v_s^2+v_A^2}\right)^{1/2} \right]^{1/2},
\end{equation}
where $v_s$ is the sound speed, $v_A$   the Alfv\'en speed, $v_c=v_Av_s/(v_A^2+v_s^2)^{1/2}$   the cusp speed, and $\theta$   the angle between the propagation direction and the magnetic field. The positive (negative) sign is for the calculation of the phase speed of the fast (slow) wave. In a plasma where $v_s > v_A$ (approximately $\beta > 1$), the phase speed of fast waves takes the shape of a flattened quasi-circle with $v_\mathrm{fa}(\theta=0,\pi)=v_s$ in the magnetic field direction and $v_\mathrm{fa}(\theta=\pi/2,3\pi/2)=(v_A^2+v_s^2)^{1/2}$ perpendicular to it. On the other hand, slow waves take the shape of double quasicircles  with $v_\mathrm{sl}(\theta=0,\pi)=v_A$ in the magnetic field direction and $v_\mathrm{sl}(\theta=\pi/2,3\pi/2)=0$ perpendicular to the magnetic field. Therefore, also for slow waves there is still a non-zero phase speed for all directions except exactly perpendicular to the magnetic field. This effect was previously used by \citet{nakariakov_zimovets_2011} to explain flare ribbon propagation.

Assuming local homogeneity and utilizing Equation \ref{eq:phasespeed}, we can apply the Huygens-Fresnel principle to theoretically predict the locations of fast and slow wave fronts. In order to do this we assume that the wave originates from a point source. In this point the phase speed in all directions is calculated, supplying us with information of the wave front location in the next snapshot. For all subsequent snapshots we calculate the phase speed in each point of the previous wave front. The next fast (slow) wave front is then the outer edge of all fast-wave quasicircles (slow-wave double quasicircles).

\begin{figure*}
  \centering
  \includegraphics[width=0.45\textwidth]{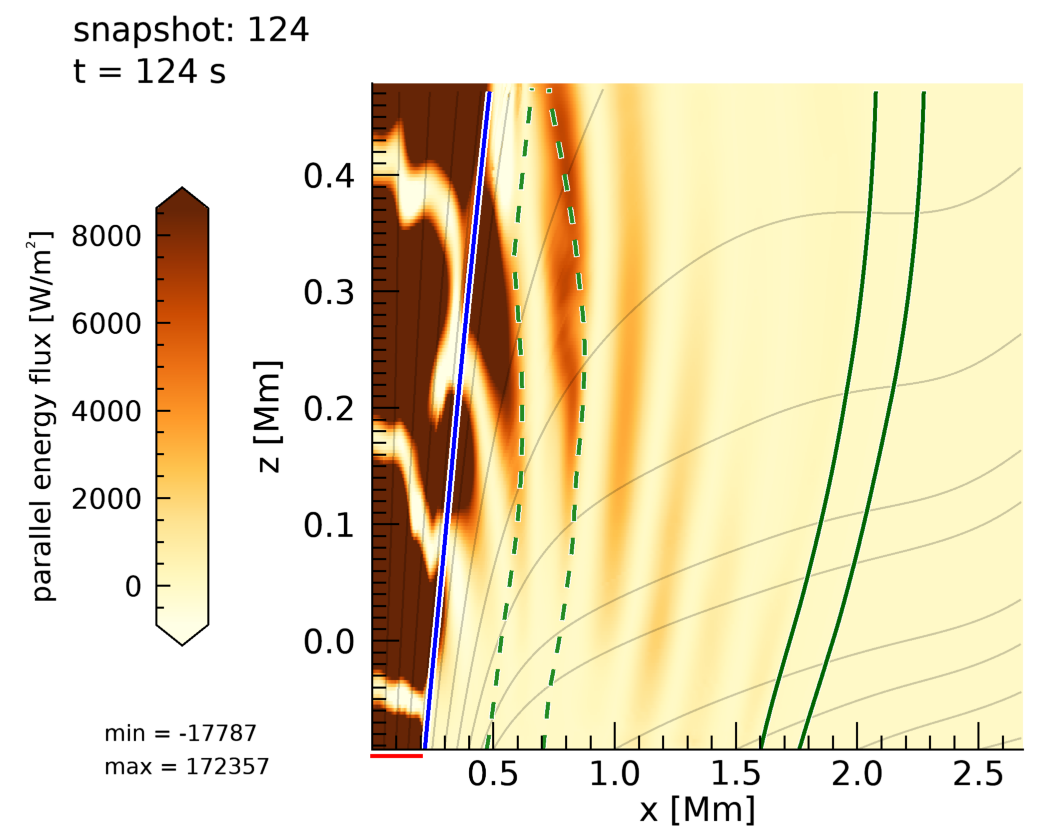} \includegraphics[width=0.45\textwidth]{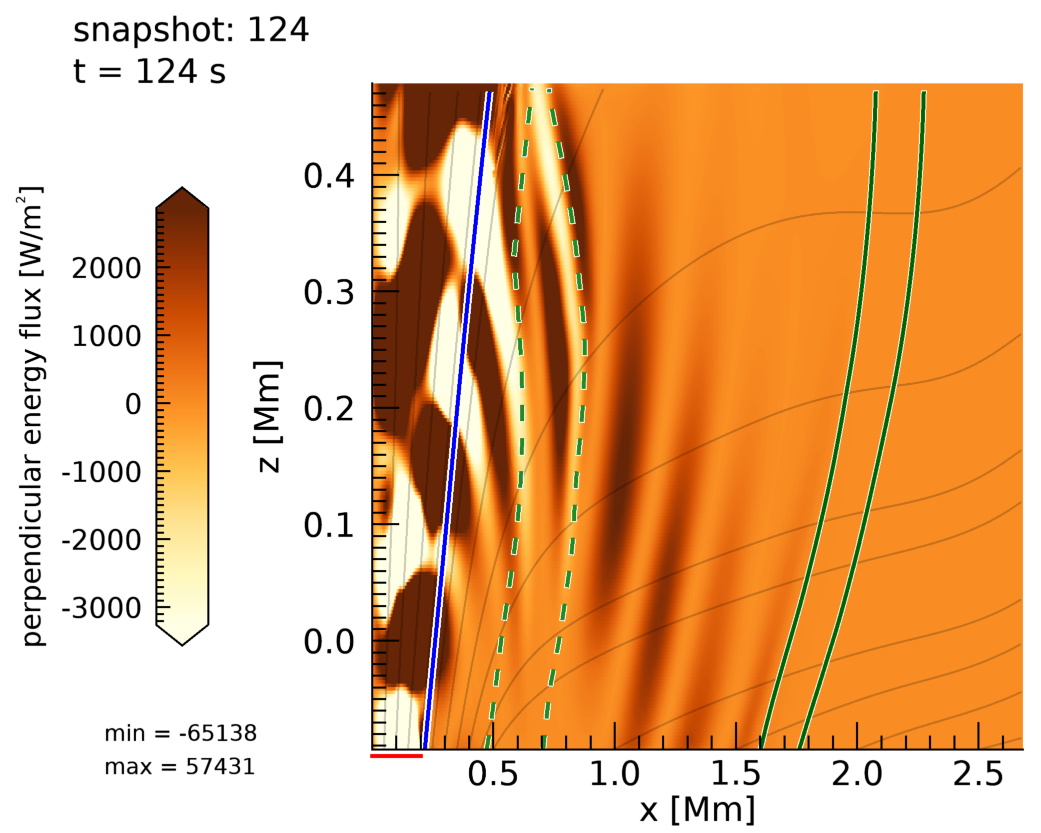}
  \caption{Snapshot of the wave energy flux parallel (left) and perpendicular (right) with respect to the magnetic field. The color range is saturated. The solid (dashed) green lines show the first theoretical wave fronts of the fast (slow) waves; the gray lines show the magnetic field lines. The red bars below the $x$-axis indicate the driver location. The blue lines highlight the field line considered for the analysis in Figure \ref{fig:flux_slab_driver}. Movies of the full time sequence are available online.}
  \label{fig:flux_snapshot_slab}
\end{figure*}

We let our theoretical wave fronts for both fast and slow waves propagate from two point sources at the bottom of the domain: one at the pore axis at $(x,z)=(0,-0.095)$ Mm and one at the edge of the driver at $(x,z)=(0.2,-0.095)$ Mm, starting from $t=0$ s. Figure \ref{fig:flux_snapshot_slab} shows the saturated wave energy flux parallel (left) and perpendicular (right) to the magnetic field at $t=124$ s. The theoretical fast wave fronts (solid lines) and slow wave fronts (dashed lines) are overplotted in green. By examining the time sequence, which is available as movies online, it is evident that there are many waves propagating with the exact same shape and speed as the theoretical fast wave fronts outside the pore. We therefore identify those waves as fast waves. They can be  seen most clearly  in Figure \ref{fig:flux_snapshot_slab} between $x\approx 0.7$ Mm and $x\approx 1.7$ Mm.

There are also waves propagating out of the pore with the same shape and speed as the theoretical slow wave fronts. We therefore identify these waves as slow waves. In Figure \ref{fig:flux_snapshot_slab} they can be predominantly seen in the perpendicular flux component between the two dashed lines. When observing the time sequence for the parallel component and focusing on that region an interaction between fast and slow waves can be seen. However, at this point the flux has already exited the pore, and we  thus do not discuss this further.

When observing the full time sequence of the movies of Figure \ref{fig:flux_snapshot_slab}, the theoretical wave fronts eventually  develop a dip close to the border of the pore (e.g., snapshot 35). This is especially prominent for the fast waves, and  is also seen in the simulation data. The reason for this dip is the density structure at that location, which can be seen in Figure \ref{fig:density}. The difference in density leads to a difference in phase speed.

Although there are clearly waves leaking out of the pore, most of the flux is contained within the pore, following the magnetic field lines. To estimate the effect of lateral leakage on the damping of energy flux with height, we compare the time-integrated total flux present along the field line highlighted in blue in Figures \ref{fig:vz_slab_driver} and \ref{fig:flux_snapshot_slab} (which is the total flux lost laterally) with the time-integrated total flux inside the pore at the bottom of the domain (which is the total incoming flux). The time integration of the flux is calculated for the first wave front over one period $T$ for all locations
\begin{equation} \label{eq:time_integration}
    \Vec{E_t}(x,z)=\int_{t_1(z)}^{t_2(z)} \Vec{E}(x,z,t) dt,
\end{equation}
where $\Vec{E}$ is the wave energy flux according to Equation \ref{eq:flux_basic} and $t_1(z)$ and $t_2(z)$ is  the time of the beginning and  end, respectively, of the first wave front at height $z$, with $t_2(z)=t_1(z)+T$.

For the calculation of the escaped flux we chose a field line rooted slightly outside the driver region in order to be sure that all flux at that location has exited the pore. We then integrate the time-integrated flux components along this field line from the root of the field line at the bottom of the domain until height $z$, before calculating the time-integrated total flux. The total escaped flux is then
\begin{equation}
  E_{t,\mathrm{esc}}(z)=\left( \left[ \int_0^{l(z)} E_{t,\parallel}(x,z) dl \right]^2 +  \left[ \int_0^{l(z)} E_{t,\perp}(x,z) dl \right]^2 \right)^{1/2},
\end{equation}
where $l(z)$ is the length of the field line at height $z$ and the integrals of the fluxes are taken along the field line. Here $E_{t,\parallel}$ and $E_{t,\perp}$ are the parallel and perpendicular components of the time-integrated energy flux (Equation \ref{eq:time_integration}) with respect to the magnetic field (and therefore the field line), with $E_{t,\parallel} \perp E_{t,\perp}$. The integration is done before calculating the absolute value to allow flux with opposing signs to cancel out. 

Similarly, the total flux contained in the pore at the bottom of the domain is calculated by
\begin{equation}
  E_{t,\mathrm{bot}}=\left( \left[ \int_0^{x_l} E_{t,x}(x,z=z_\mathrm{bot}) dx \right]^2 +  \left[ \int_0^{x_l} E_{t,z}(x,z=z_\mathrm{bot}) dx \right]^2 \right)^{1/2},
\end{equation}
where $x_l=0.22$ Mm is the $x$-position of the field line root, $z_\mathrm{bot}=-0.095$ Mm is the $z$-location of the bottom of the domain, and the integrals are taken horizontally across the pore at the bottom of the domain. Here $E_{t,x}$ and $E_{t,z}$ are the $x$- and $z$-components of the time-integrated energy flux, with $E_{t,x} \perp E_{t,z}$.

The effect of wave leakage on the damping is then estimated by
\begin{equation} \label{eq:damping}
    e_\mathrm{damp}(z)=1-\frac{E_{t,\mathrm{esc}}(z)}{E_{t,\mathrm{bot}}}.
\end{equation}
The result of Equation \ref{eq:damping} is shown in Figure \ref{fig:flux_slab_driver} (right) as the blue dashed line. There is a significant difference between this line and the line showing missing flux when only considering geometric spreading (orange dash-dotted line). Both methods are   estimates, and we expect the actual effect of lateral wave leakage to lie between these lines.


\section{Conclusions and  discussion} \label{sec:discussion}

We created a MHS model close to equilibrium, which was inspired by observational data of a solar pore \citepalias{gilchrist_etal_2021} and investigated possible damping mechanisms by driving the model with a vertical velocity perturbation at the bottom of the domain. We found that, even if viscosity, resistivity, or thermal conduction are included, the strong damping from the observations could not be reproduced at all by using a driver that covers the whole bottom boundary. When switching to a localized driver, however, the results show strong damping in our simulations. This damping occurs because of a) geometric spreading, where the flux is spread over a wider area due to diverging field lines and b) lateral wave leakage, where waves leave the pore. Therefore, even if only considering classic wave effects, significant damping can be achieved. Wave leakage at the edge of a solar pore was indeed already observed by \citet{stangalini_etal_2011}.

\subsection{Effects of differences between observed pore and model \& comparison of simulations to observations} \label{subsec:differences_obs_model}

It was mentioned in Section \ref{subsec:model} there are differences between our model and the observational data example pore \citepalias[][pore 3]{gilchrist_etal_2021}. The differences in density and pressure profiles mainly lead to differences in characteristic wave speeds. This does not affect the damping due to geometric spreading, as this damping mechanism is only dependent on the magnetic field structure, which is similar to the observations, with nearly vertical inclination inside the pore and nearly horizontal field lines outside. 

\begin{figure}
  \centering
  \includegraphics[width=0.45\textwidth]{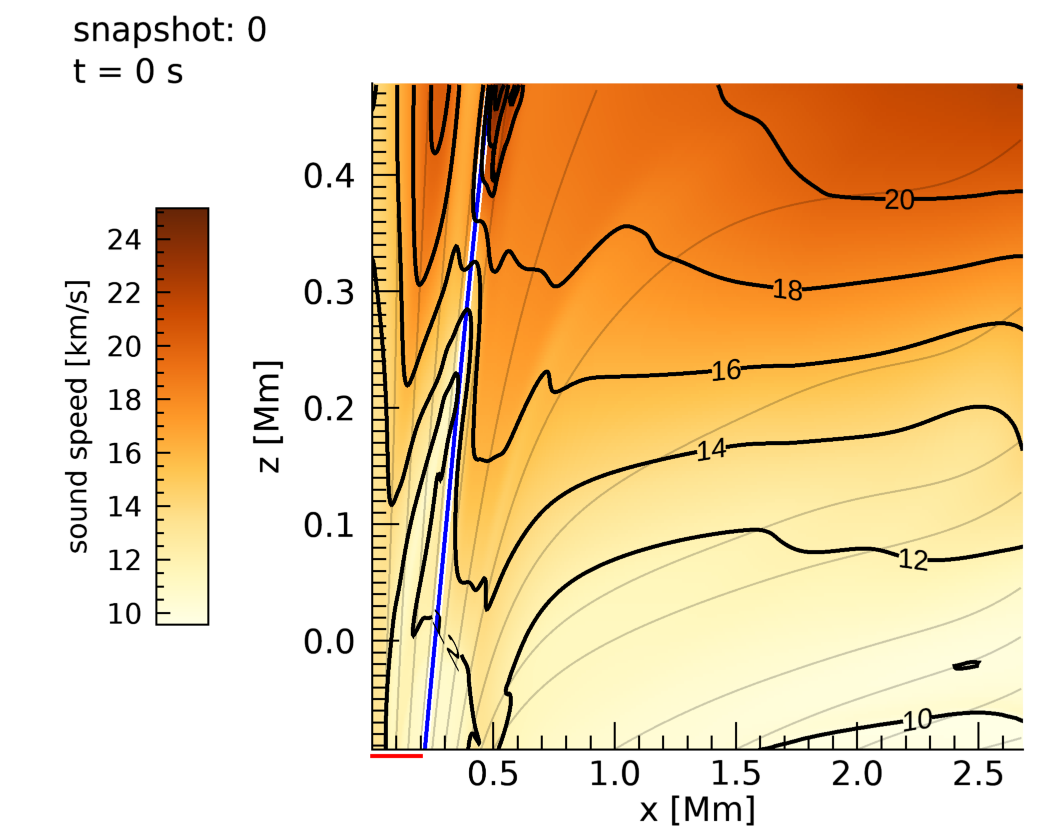} 
  \caption{Sound speed of the initial atmosphere. Gray lines show magnetic field lines. The red bar below the $x$-axis indicates the driver location for simulations with localized driver. The blue line highlights the field line considered for the analysis in Figure \ref{fig:flux_slab_driver}. Contours for the sound speed are shown in thick black lines.}
  \label{fig:cs0}
\end{figure}

An important point we have to note, however, is the sound speed profile, as shown in Figure \ref{fig:cs0}. In our model, the sound speed generally increases with height, whereas it is the opposite for the observations. In addition, there is a strong horizontal structuring, with lower speeds at the center and the border of the pore. From applying Snell's refraction law, as also discussed in the context of sunspots by \citet{khomenko_collados_2006}, we know that waves travelling into a medium with higher phase speed refract away from the line perpendicular to the constant-phase-speed-line. If in our simulations the fast (acoustic) waves are propagating along the diverging field lines, they are refracted away from the pore. Therefore, should the fast lateral waves in our simulations exclusively occur because of refraction, we would not expect acoustic waves escaping laterally for the observations of pores like in \citetalias{gilchrist_etal_2021}. The effect of lateral leaking for magnetic waves should be the same, however, as the Alfv\'{e}n speed profile in our simulations is similar to the observations.

Evidence of at least some fast wave refraction occurring in our simulations is seen in the amplitude of the wave energy flux, where the amplitude is increased at the center and the boundary of the pore compared to the region in between. Those regions coincide with regions of lower sound speeds and waves are therefore refracted toward those regions. The increased amplitude in the pore boundary therefore does not occur because of sausage surface waves. Since the sound speed is higher at the location just outside the pore, waves that are located outside the pore would be refracted into the pore. This could be one of the reasons why the energy flux profile increases with height for the full driver (Figure \ref{fig:flux_full_driver}), as there are ample waves present outside the pore to be refracted. In addition, fast wave energy flux that escaped from the pore was eventually refracted down toward the bottom of the domain in the simulations with localized driver. This can be seen in Figure \ref{fig:flux_snapshot_slab} (right), where the perpendicular flux component for the fast waves outside the pore is mainly positive and therefore directed downward, considering the nearly horizontal field lines. This refraction of fast waves is similar to what was found by \citet{khomenko_collados_2006}. 

The observations of pore 3 \citetalias{gilchrist_etal_2021} also show higher energy flux concentrations at the pore boundaries. Contrary to the events in our simulations, it was found that these flux concentrations are due to surface sausage modes. This could possibly promote additional lateral wave leakage as flux already present at the edge of the pore could more easily escape.

A crucial difference between observations and simulations is that due to the cadence of the instruments, \citetalias{gilchrist_etal_2021} were only able to investigate slow waves, whereas in this paper we have a combination of slow and fast waves. By splitting the energy flux into magnetic (Poynting) and hydrostatic contributions, slow and fast waves could have been studied separately. However, most of the slow waves in our simulations with localized driver stem from the sharp edge of the step-function, causing most of the slow waves being concentrated just inside and atop the considered field line marking the boundary of the pore in our analysis (blue highlighted field line in e.g. Figure \ref{fig:flux_snapshot_slab}), with little  slow wave flux inside the rest of the pore. We therefore only considered the total flux for our analysis, as our estimate for the influence of damping due to lateral wave leakage (Equation \ref{eq:damping}) would not have worked for slow waves alone. On the other hand, there was no need to exclude slow waves from the same analysis as the magnitude of the Poynting flux is about three orders of magnitude smaller than the hydrostatic component. 

In our model $\beta > 1$ everywhere, whereas $\beta < 1$ is expected inside the pores according to the observations. This basically means that the fast waves inside the pore in this paper correspond to the slow waves observed in \citetalias{gilchrist_etal_2021} as they both have predominantly acoustic properties. While slow waves are allowed to propagate in all directions except directly perpendicular to the magnetic field (see Equation \ref{eq:phasespeed}), their phase speed as a function of angle to the magnetic field  has a different shape than for fast waves. While, according to our results, slow waves also leave the pore, it is possible that due to this different shape fewer low $\beta$ slow acoustic waves (observations) would leave the pore than fast acoustic waves in our simulations. However, the slow acoustic waves in observations are still comparable to the fast acoustic waves simulated here. It is therefore reasonable to assume that within the pore the slow wave energy flux would be dominant over the fast wave energy flux if our model atmosphere had $\beta < 1$ in that region. Applying this assumption to the real world highlights one of the difficulties in observing fast modes: the fast wave flux would be overshadowed by the slow wave flux. In addition, having a low plasma-$\beta$ inside the pore inevitably leads to a $\beta=1$ (or $v_s=v_A$) layer at the border of the pore with high $\beta$ outside. In these layers waves are strongly subjected to mode conversion \citep{cally_2005,cally_2006,schunker_cally_2006,hansen_etal_2015}. Whether these mode conversions   increase the amount of energy flux escaping from the pore or   have a channeling effect in the pore will have to be determined in future work.

\subsection{On other limitations of the current study} \label{subsec:limitations}

In this work, we did not account for any radiative losses. According to \citet{carlsson_stein_2002}, acoustic waves in the photosphere are much more damped at higher frequencies, meaning that the impact of this damping mechanism in our simulations would be larger than for the observations of \citetalias{gilchrist_etal_2021}, who observe longer periods.

Our simulations were done on a 2D Cartesian grid. In 2D, the ``area'' inside the pore at each height is just a 1D line. Therefore, we estimated the damping due to geometric spreading to be proportional to $1/R(z)$ with $R(z)$ the distance between the pore axis and a field line. In 3D, however, we   expect the wave energy flux due to this effect to decrease with $1/R(z)^2$. Estimating the change in effect from 2D to 3D for wave leakage is more difficult. We assume that it is dependent on the ratio of the area inside the pore to the area that has been available for flux to escape, which is the mantle of the pore up to a specific height. This ratio is $R(z)/l(z)$ in 2D and $R(z)^2\pi/(2R(z)\pi l(z))$ in 3D, with $l(z)$ describing the length of the considered field line from the root up to a certain height $z$. Therefore, the dependence $R(z)/l(z)$ can also be assumed for 3D. The increase in efficiency of geometric spreading for 3D could account for the difference between the damping in our simulations and the observed damping. We note that by assuming a 2D geometry in our simulations we have excluded the possibility of Alfv\'{e}n waves.

\subsection{Concluding remarks and future work} \label{subsec:concluding_remarks_future}

As discussed above, there are both slow and fast waves present in our simulations. The slow waves are predominantly excited at the edge of the step-function driver. Simulations using a Gaussian-shaped driver instead show that slow waves are excited at the flank of the Gaussian, mostly at the steepest location. This leads to the conclusion that any kind of localized vertical driver would excite both slow and fast waves. Therefore, we   also expect both kinds of waves to be present in the photosphere at all times. While slow modes have been observed in the photosphere many times, temporal resolution has so far   limited similar studies for fast waves. However, future instruments on the Daniel K. Inouye Solar Telescope (DKIST), European Solar Telescope (EST), and National Large Solar Telescope (NLST) might provide the cadence needed to observe fast waves propagating at an inclined angle with respect to the magnetic field.

Observing the leaking waves as seen in our simulations might be challenging as the magnitude of the vertical (line-of-sight) velocity perturbations is roughly a factor of ten lower than the perturbations inside the pore. However, since the wave fronts of the leaking waves are inclined from the vertical (as seen in Figure \ref{fig:flux_snapshot_slab}), an observer from above would see the integrated effects of waves in different phases (i.e., positive and negative velocities within the same pixel). This would lead to spectral line broadening. The possibility to observe the leaking waves using this effect can be investigated using forward modeling techniques, such as the FoMo code developed by \citet{vandoorsselaere_etal_2016_FOMO}.


\begin{acknowledgements}
      The authors thank the referee for their constructive comments. JMR and TVD have received funding from the European Research Council (ERC) under the European Union's Horizon 2020 research and innovation programme (grant agreement No. 724326). CAG-M, DBJ, and SDTG are grateful to Invest NI and Randox Laboratories Ltd. for the award of a Research \& Development Grant (059RDEN-1) that allowed the research framework employed to be developed. DBJ and SDTG also wish to acknowledge the UK Science and Technology Facilities Council (STFC) for funding under the Consolidated Grant ST/T00021X/1.
\end{acknowledgements}

\bibliographystyle{aa}
\bibliography{main}

\begin{appendix}

\section{Simulations with driver below the cutoff frequency} \label{appendix:7_min}

In order to focus on propagating waves all simulations in this paper have so far   been conducted with a driver frequency well above the expected cutoff frequency of the model atmosphere, and therefore also with a period much smaller than the five-minute waves observed by \citetalias{gilchrist_etal_2021}. This means that for the current study, all effects of the cutoff frequency have been ignored. However, as mentioned in Section \ref{sec:introduction}, even if the waves of \citetalias{gilchrist_etal_2021} definitely have a propagating character, they might be altered to partly evanescent waves by the existence of the cutoff frequency \citep{centeno_etal_2006}. In this section we  explore the possibility of damping due to evanescent waves by conducting the same two experiments as before, namely simulations with full driver and localized driver, but with a lower driver frequency.

\subsection{Cutoff frequency and new driver period} \label{subsec:appendix:cutoff}

It is commonly accepted that acoustic waves with frequencies below the cutoff frequency are not allowed to propagate, but are standing and evanescent. However, it is difficult to define an exact value for the cutoff frequency, and numerous different definitions exist. \citet{centeno_etal_2006} show that when radiative losses are involved there is no clear cutoff frequency that distinguishes between fully propagating or fully evanescent waves. \citet{felipe_etal_2018} compared analytical definitions for the cutoff frequency suitable for sunspot umbrae from \citet{lamb_1909}, \citet{schmitz_fleck_1998}, and \citet{roberts_2006} to the observed cutoff. The results generally agree. Using the same analytical expressions as discussed in \citet{felipe_etal_2018} on the observational data obtained by \citetalias{gilchrist_etal_2021} for pore 3 shows that waves with a period of five minutes  indeed have a lower frequency than the cutoff frequency for at least most of the observed domain.

According to the same equations, a driver period of five minutes would still result in a frequency above the cutoff frequency for our model atmosphere. To mimic the conditions of the observations, we choose a longer driver period of $T=7$ minutes for the following simulations. To include at least one full period of the driver the simulations are run for 500 seconds.

\subsection{Results}

\begin{figure}
  \centering
  \includegraphics[width=0.48\textwidth]{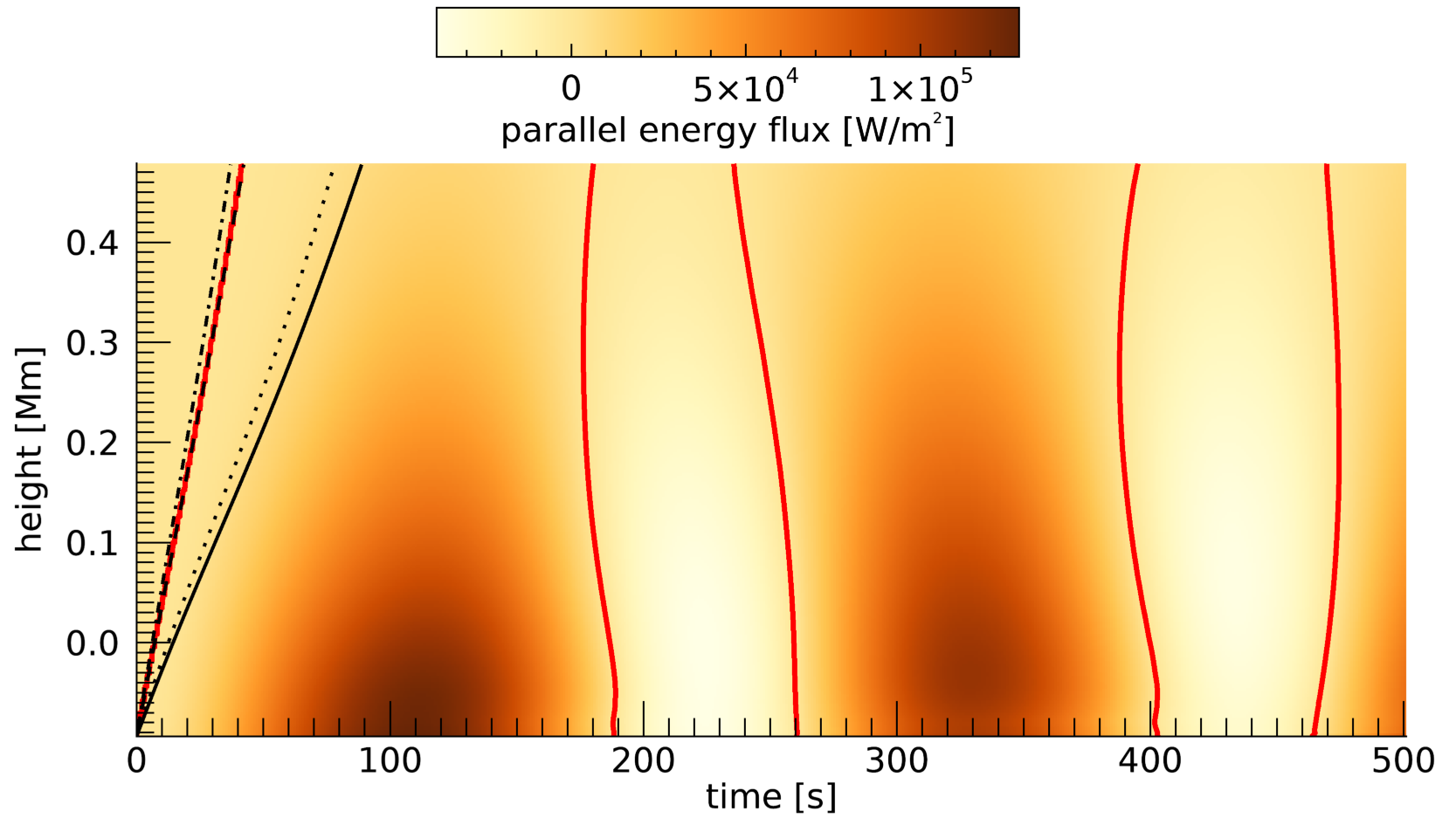}
  \caption{Parallel wave energy flux as a function of height and time at the pore axis for the simulation with localized driver with a period of 7 minutes. The black lines show (from steepest to flattest) the fast speed (dash-dotted line), sound speed (dashed line), Alfv\'{e}n speed (dotted line), and cusp speed (solid line). The red lines show the contours for zero flux. The frequency of the energy flux is approximately doubled compared to the driver period because of phase difference between $p'$ and $\Vec{v'}$ (see Equation \ref{eq:flux_basic}).}
  \label{fig:app:height-time-graph}%
\end{figure}

Figure \ref{fig:app:height-time-graph} shows the height--time graph of the wave energy flux parallel to the magnetic field at the axis of the pore for the simulation with localized driver. The characteristic speeds (starting from steepest: fast speed $v_\mathrm{fa}(\theta=\pi/2)=(v_A^2+v_s^2)^{1/2}$, sound speed, Alfv\'{e}n speed, cusp speed) are plotted as black lines, while the contour at value zero is shown in red. The initial part of the first wave (i.e., the initial disturbance where the flux is above zero for the first time) propagates with the sound speed (black dashed line overplotted on first red line) as it did for the propagating waves in Section \ref{sec:results}. Then, however, the waves get altered by the effects of the cutoff frequency to approximately standing waves within less than half a driver period, as can be seen from the nearly vertical features in the figure. This is not what was observed in \citetalias{gilchrist_etal_2021}, who found clear evidence of propagating waves. The difference might be accounted for by the neglect of radiative losses in our simulations.

\begin{figure}
  \centering
  \includegraphics[width=0.45\textwidth]{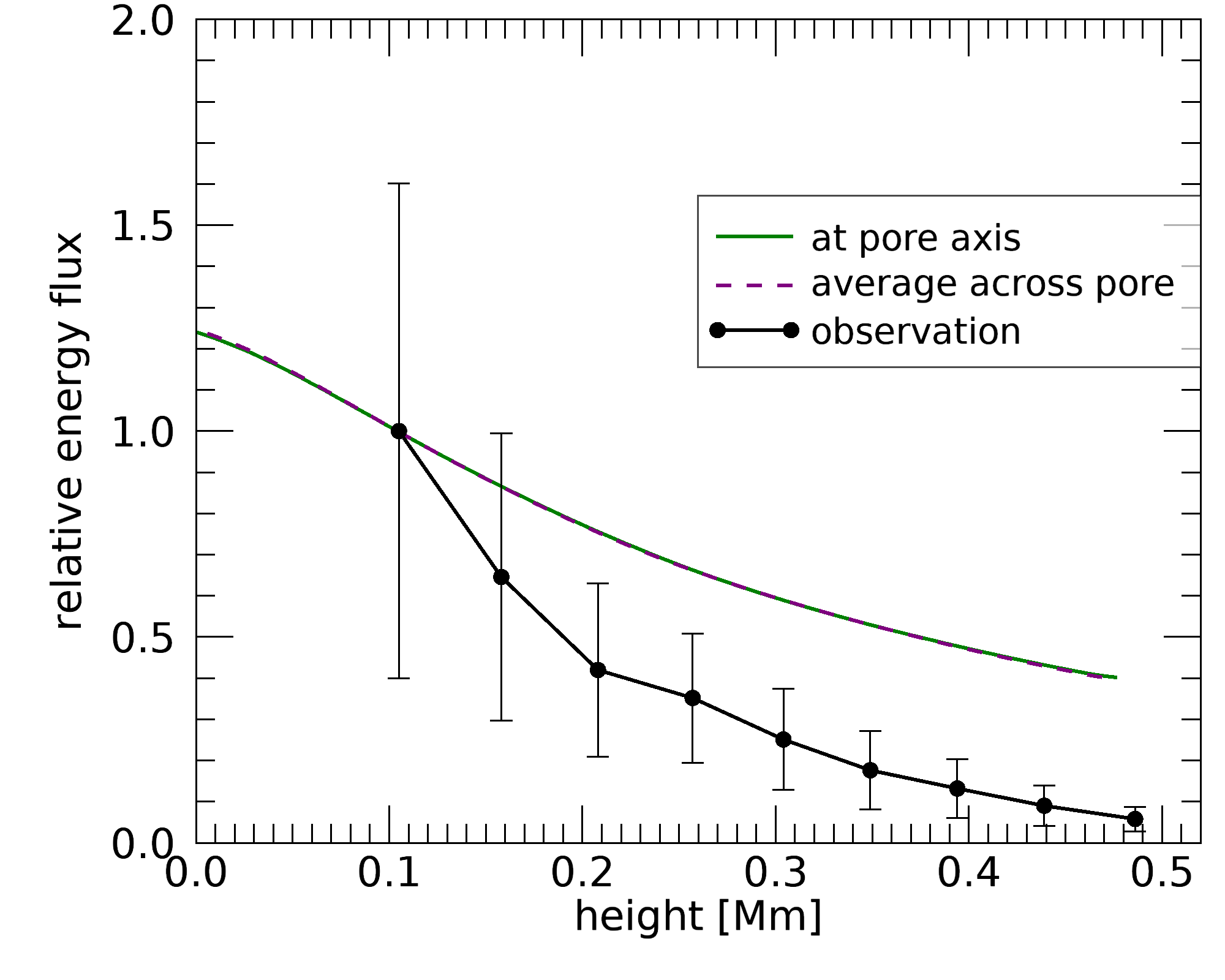}
  \caption{Same as Figure \ref{fig:flux_full_driver}, but for a driver period of 7 minutes.}
  \label{fig:app:flux_full_driver_7min}%
\end{figure}

\begin{figure}
  \centering
  \includegraphics[width=0.45\textwidth]{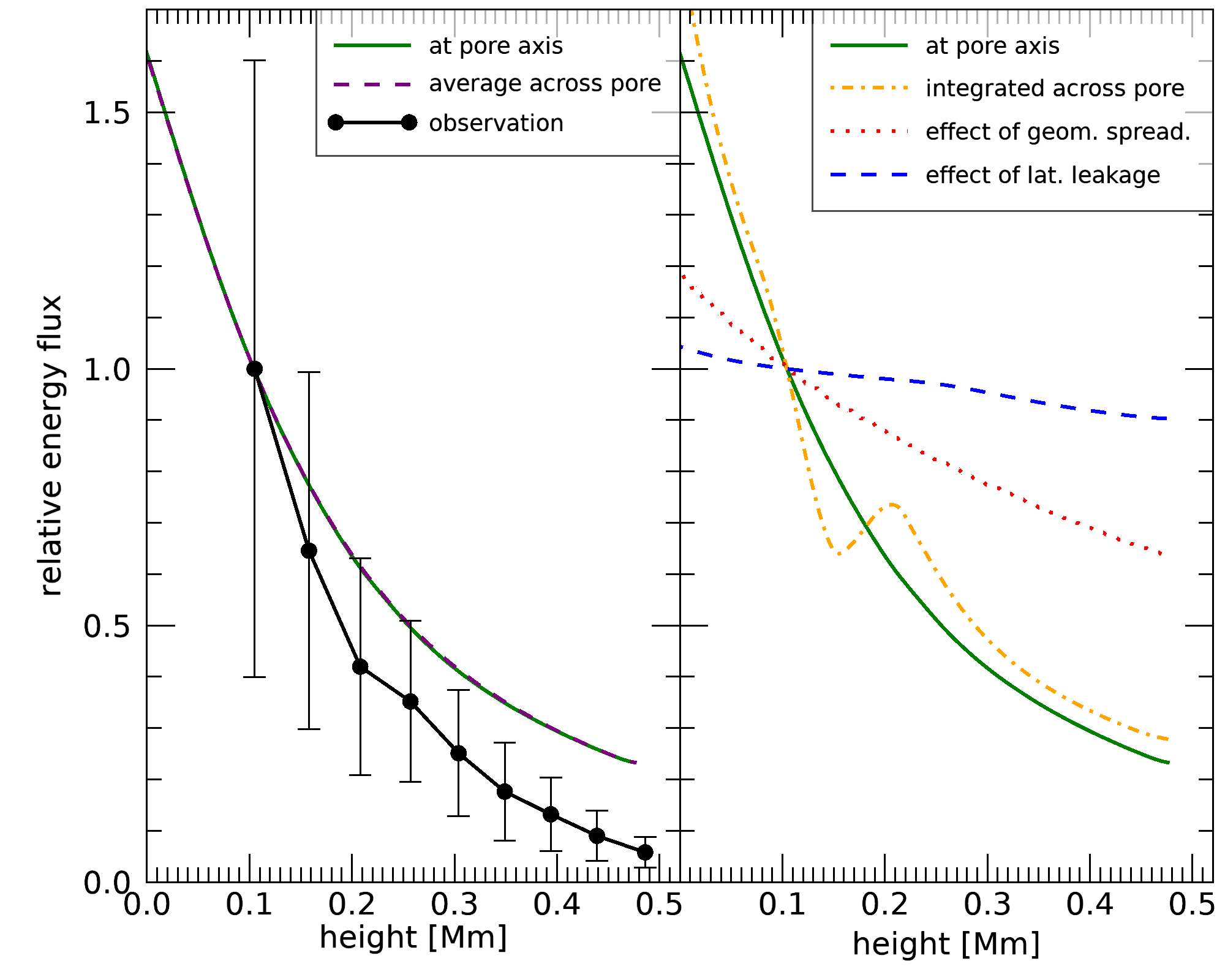}
  \caption{Same as Figure \ref{fig:flux_slab_driver}, but for a driver period of 7 minutes.}
  \label{fig:app:flux_slab_driver_7min}%
\end{figure}

We performed the same study for the wave energy flux damping as in Sections \ref{subsec:full_driver} and \ref{subsec:localized_driver}, but for the low-frequency driver. Figure \ref{fig:app:flux_full_driver_7min} shows the results for the full driver, while Figure \ref{fig:app:flux_slab_driver_7min} shows the results for the localized driver. It is immediately apparent that the energy flux for the full driver is now heavily damped as well, about the same amount as the energy flux for the localized high-frequency driver (Figure \ref{fig:flux_slab_driver}). The energy flux for the localized low-frequency driver (Figure \ref{fig:app:flux_slab_driver_7min}) is damped even more, probably because the damping with height is not decreased by inward refracted waves as for the full driver.

\subsection{Discussion}

It is obvious that the choice of driver frequency strongly affects the damping in our simulations. However, whether this is   purely due to evanescent waves is not fully clear.

On the one hand, the dash-dotted orange curve in Figure \ref{fig:app:flux_slab_driver_7min}, which shows the damping without effect of geometric spreading, strongly follows the solid green line, which is the full damping in our simulation with the localized low-frequency driver. This hints that geometric spreading has little to no effect in this case. At the same time the dashed blue line, which is an estimate for the influence of lateral leakage, is nearly constant, meaning that  this effect is also not very strong. Therefore, a crucial damping mechanism is missing, which is likely the reflection of waves due to the cutoff frequency.

On the other hand, these new simulations and their analysis are subject to some limiting factors. First of all, due to the low frequency, the wavelengths of the resulting waves are significantly longer than the size of the computational domain. This could lead to strange boundary effects influencing the results. Since the ratio of the wavelength to the size of the pore (which is smaller in our model than in the observations) also changes drastically, this could account for the decreased effects of geometric damping and lateral wave leakage. In addition, due to the waves starting at some final time $t_0$, there are no waves present in the domain before the first waves reach a certain height (i.e., left of the first red line in Figure \ref{fig:app:height-time-graph}). Therefore, when integrating the wave energy flux over time, the lower integration boundary $t_1(z)$ was chosen by using a relative threshold to determine the onset of the first wave at every height. This line basically coincides with the sound speed line (dashed) in Figure \ref{fig:app:height-time-graph}. The upper integration boundary was then determined by $t_2(z)=t_1(z)+T$, with $T$ being the driver period. Effectively, the time integration for the simulations with high-frequency driver was done over the first period of the wave, as a translation of $t_1(z)$ by $T=30$ s resulted in a $t_2(z)$ being located right in front of the next wave train. This is not the case for the low-frequency waves because they change from propagating to standing waves within the first wave period, meaning that their steepness changes in Figure \ref{fig:app:height-time-graph}. Therefore, it is not clear over which time period the integration should be performed, and the choice might affect the shape of the damping curves in Figures \ref{fig:app:flux_full_driver_7min} and \ref{fig:app:flux_slab_driver_7min}.

Moreover, even if the limitations listed above have little to no effect, there are still no propagating waves in our low-frequency simulations, as opposed to the observations of \citetalias{gilchrist_etal_2021}. Therefore, the damping in the low-frequency simulations due to evanescent waves is expected to be much stronger than for the observations, where the waves were at least partly propagating. This validates the study of the other damping mechanisms presented in this paper.

\end{appendix}

\end{document}